\documentclass[12pt]{iopart}
\pdfoutput=1
\usepackage[usenames,dvipsnames]{color}
\usepackage{graphicx}
\usepackage[utf8]{inputenc}
\usepackage[T1]{fontenc}
\usepackage{iopams}

\begin{document}
\title{Energetics of a driven Brownian harmonic oscillator}

\author{Mohammad Yaghoubi}
\address{Complexity Science Group, Department of Physics and Astronomy, University of Calgary, Calgary,
Alberta T2N 1N4, Canada}

\author{M. Ebrahim Foulaadvand}
\address{Department of Physics, University of Zanjan, P.O.Box 45195-313, Zanjan, Iran}
\ead{foolad@znu.ac.ir}

\author{Antoine B\'{e}rut}
\address{Aix-Marseille Univ, CNRS, IUSTI, Marseille, France}

\author{Jerzy {\L}uczka}
\address{Institute of Physics, University of Silesia, 40-007 Katowice, Poland}
\address{Silesian Center for Education and Interdisciplinary Research, University of Silesia, 41-500 Chorz{\'o}w, Poland}
\ead{jerzy.luczka@us.edu.pl}

%
\begin{abstract}
 {\it Abstract.}  We provide insights into energetics of a Brownian oscillator in contact with a heat bath and driven by an external unbiased time-periodic force that takes the system out of thermodynamic equilibrium.  Solving the corresponding  Langevin equation, we compute average kinetic and potential energies in the long-time stationary state.   We also derive the energy balance equation and study the energy flow in the system. In particular, we identify the energy delivered by the external force, the energy dissipated by a thermal bath  and the energy provided by thermal equilibrium fluctuations. Next, we illustrate Jarzynski work-fluctuation relation and consider the stationary state fluctuation theorem for the total work done on the system by the  external force.  
 Finally, by determining time scales in the system, we analyze the strong damping regime  and discuss the problem of overdamped dynamics when  inertial effects can be neglected.
\end{abstract}
\pacs{
05.10.Gg, 
05.40.-a, 
05.40.Ca, 
05.60.-k, 
}
\maketitle

\section{Introduction}

Classical thermodynamics and statistical physics usually deals with systems either isolated or at equilibrium with their environment. For example, if a system is in contact and weakly interacts with a thermostat (heat bath), its stationary state is a  Gibbs canonical equilibrium state~\cite{Reif}. However, this ``ideal'' situation is not always satisfied and real systems often have strong couplings with their environments or are driven far from thermodynamic equilibrium by an external agent. Recently, a lot of theoretical developments have improved our understanding of non-equilibrium thermodynamics  such as fluctuation theorems
\cite{BK77,ECM93,GC95,J97,C99}, which  have been generalized
 both in  classical and quantum statistical physics \cite{BK77,J11,EHM09,CHT11,HT15}. Moreover, a number of important results have been obtained in the theory of stochastic thermodynamics ~\cite{Sekimoto2,Seifert,van}.
These approaches are particularly relevant for  small systems where thermal fluctuations cannot be neglected, such as colloidal particles, polymers, molecular motors, etc. \cite{lowen}. Far from thermal equilibrium systems exhibit many fascinating and non-intuitive properties like absolute negative mobility \cite{ANM}, stochastic resonance \cite{SR}, non-monotonic temperature dependence of a diffusion coefficient \cite{DT} or transient but long-lasting anomalous diffusion \cite{trans,trans2}.
Interestingly, in this recent framework, the Langevin equation, first introduced in 1908 to describe the Brownian motion~\cite{Langevin}, has become a paradigmatic study case.

The intention of this paper is to present in a pedagogical way an  example  of a non-equilibrium system. It should be useful for students in statistical physics to follow  step by step a derivation of some formulas and next to perform  detailed analysis of the system properties, in particular its energetics.  We consider only one model: the classical harmonic oscillator in contact with a heat bath and driven by an external, unbiased  time-periodic force. The system is described by a Langevin equation. It is simple enough so that the main statistical quantities characterizing its dynamics can be calculated exactly in analytical forms. Moreover, it is a relevant model for a lot of experimental applications at small scales. For example, optical tweezers used to trap and manipulate micrometric objects are very well described by a harmonic potential for small displacements~\cite{Simmons,Tlusty}. Nearly forty years after their invention~\cite{Ashkin}, optical tweezers have become a very common tool and are now widely used in biology and physics laboratories~\cite{Grier,booktweezers1,booktweezers2}. Another classical example of applications is a  small torsion pendulum~\cite{exp1,exp2,ForceGauss}. It has also been shown in a one dimensional system of hard particles that the hard core interaction among particles can lead to a restoring force that can be modelled by a harmonic potential~\cite{foolad1}.  Moreover, a two-dimensional diffusion of a Brownian particle in a periodic asymmetric channel with soft walls can been modeled by a parabolic potential which mimics a restoring force~\cite{Artem1,Artem2}. Harmonic potential has been also used in modelling confined colloidal systems ~\cite{Lowen09}.  More generally, any system that undergoes thermal fluctuations close enough to its stable equilibrium position can be described as a Brownian harmonic oscillator.

In this article, we have chosen to consider the case of an external sinusoidal  force acting on the oscillator  and we focus on  interpretation of the oscillator energetics. Namely, we analyze the long time dynamics and compute the potential and kinetic energies, and we describe the energy balance between the system, the thermal bath and the external force. We also discuss in details the regime  of strong damping (frequently named overdamped) dynamics where the inertia term can be neglected. We show that the overdamped regime can be rigorously analyzed by revealing  characteristic time scales in the system and then by  transforming  the Langevin equation to its dimensionless form. The similar procedure can be applied to energetics of the system.

The article is organized as follows: In section 2 we describe the model of a periodically driven Brownian particle in a harmonic potential  and we present the Langevin equation that governs its dynamics. We also define three kinds of  averages of dynamical variables of the oscillator. In section 3, we present a solution of the Langevin equation for the position and the velocity of the particle.  Next, in section 4, we compute the averaged kinetic and potential energies of the system in the long-time regime. In section 5 we derive equations for the energy balance and we discuss the energy exchanges between the system, the heat bath, and the external driving force. In section 6, we analyse period-averaged energy in dependence on the frequency of the external driving. 
In section 7,  we study the Jarzynski-type relation for the work performed by the external time-periodic force. In section 8 , 
we discuss the regime of strong damping and discuss the problem of overdamped dynamics.  Finally, in  section 9 we conclude the paper with  discussion and  summary of the findings. Appendices contain supplementary materials on Green functions (Appendix A), the particle position and velocity mean square deviation (Appendix B) and the first two statistical moments  of the work (Appendix C).   

\section{The forced Brownian harmonic oscillator}

\subsection{Langevin equation}

We consider a classical Brownian particle of mass $m$ in contact with a thermal bath of temperature $T$. For simplicity we consider its motion in  one dimension in the harmonic potential $U(x)=k x^2/2, \,k>0$. It is driven by an unbiased, symmetric  time-periodic force $F(t)=F_0\cos(\omega t)$ of amplitude strength $F_0$ and angular frequency $\omega$.  The dynamics of the particle is described by a Langevin equation in the form~\cite{chaos}:

\begin{equation}
\label{Lan}
m\ddot x + \gamma \dot x +kx=  F(t)+ \sqrt{2\gamma k_B T}\,\xi(t)~~~t \geq 0.
\end{equation}
%
Here, $x=x(t)$ is the particle coordinate, the dot denotes differentiation with respect to the time $t$ and $\gamma$ is  the friction coefficient. Usually $\gamma$ is the Stokes coefficient: $\gamma = 6 \pi R \eta$, with $R$ the radius of the particle and $\eta$ the viscosity of the surrounding fluid. However, in this paper we do not refer to particle's geometry and will simply take it as point-like one. The restoring potential force is 
\mbox{$-d U(x)/dx =-kx$}. Thermal fluctuations due to coupling of the particle with the thermal bath are modeled by a $\delta$-correlated Gaussian white noise $\xi(t)$ of zero mean and unit intensity:
\begin{equation}
\langle \xi(t) \rangle = 0, \quad \langle \xi(t)\xi(u) \rangle = \delta(t-u).
\label{xi}
\end{equation}
The noise intensity factor $2\gamma k_B T$ (where $k_B$ is the Boltzmann constant) comes from the fluctuation-dissipation theorem~\cite{CW51,kubo,zwanzig}. It ensures that the system stationary state is the canonical Gibbs state (thermodynamic equilibrium) under the dynamics governed by equation (\ref{Lan}) for vanishing driving, i.e., when $F_0 = 0$.\bigskip

\noindent The following points are noteworthy:
\begin{enumerate}
\item The system can be seen as a Brownian particle moving in a harmonic potential $U(x)$ or equivalently as a harmonic oscillator.
\item The stochastic process $x(t)$ determined by equation (\ref{Lan})  is {\it non-Markovian}. However, the pair $\{x(t), \dot x(t)\}$ constitutes a Markovian process.
\item Due to the presence of the external time-periodic force $F(t)$ the system tends for long time to a unique asymptotic {\it non-equilibrium} state characterized by a temporally periodic probability density~\cite{jung}. In this long-time state, the average values of dynamical variables are periodic functions of time, that have the same period as the acting force $F(t)$.
\end{enumerate}

\subsection{Notations for averages}

\noindent In this problem, we define three kinds of averages:
\begin{itemize}
\item We denote by $\langle Y(t) \rangle$ the ensemble averaging of a dynamical variable $Y(t)= Y[x(t), \dot{x}(t)]$ over all realizations of the Gaussian white noise $\xi(t)$ and over all initial conditions $\{x(0), \dot{x}(0)\} $.
\item We denote by $\langle Y(t) \rangle_{st}$ the average value in the long-time limit state where all transient effects die out:
\begin{equation}
\langle Y(t) \rangle_{st} =\lim_{\gamma_0 t >> 1}\; \langle Y(t) \rangle. 
\end{equation}
In this state, the dynamics is characterized by a temporally periodic probability density  with the same period $\mathsf{T}$  as the acting force: $\mathsf{T} =2\pi / \omega$.
\item We denote by $\overline{Y}$ the time averaging of $\langle Y(t) \rangle$ over one period $\mathsf{T} =  2\pi / \omega$ in the long-time regime:
\begin{equation}
\label{averY}
\overline{Y}= \lim_{t\to\infty} \frac{1}{\mathsf{T}}
\int_{t}^{t+\mathsf{T} } \langle Y(u) \rangle \; du. 
\end{equation}
By doing this, we obtain the time-independent asymptotic characteristic average of the dynamical variable $Y(t)$.
\end{itemize}
In the following sections, we will use the three notations: $\langle Y(t) \rangle$, $\langle Y(t) \rangle_{st}$ and $\overline{Y}$.

\section{Solution of the Langevin equation}

\subsection{Position of the Brownian particle }

The Langevin equation (1) is a stochastic non-homogeneous linear differential equation and its solution 
$x(t)$ is in the form  $x(t) = x_c(t) + x_p(t)$, where the complementary part $x_c(t)$ satisfies the homogeneous differential equation (i.e., when $F_0=0$ and $\xi(t)=0$), with the initial condition $x_c(0)=x_0$ and $\dot{x}_c(0)=v_0$. The particular solution $x_p(t)$ satisfies the inhomogeneous equation with the initial conditions $x_p(0)=0$ and $\dot{x}_p(0)=0$.   It is convenient to decompose the particular solution $x_p(t)$ into two additive parts: $x_p(t)=x_d(t) + x_{\xi}(t)$, where $x_d(t)$ is a particular solution corresponding to the deterministic force $F(t)$ and $x_{\xi}(t)$ is a solution related to the noise term $\xi(t)$. We therefore write 
\begin{equation}
x(t) = x_c(t) + x_d(t) + x_{\xi}(t).
\label{x(t)}
\end{equation}
The term $x_c(t)$ is a linear combination of two independent solutions of the homogeneous  equation (in the case when $F_0=0$ and  $\xi(t)=0$):
\begin{eqnarray}
	x_c(t)= A \mbox{e}^{-\gamma_0 t} \sin (\Omega t) + B
	 \mbox{e}^{-\gamma_0 t} \cos (\Omega t),
\label{x0}
\end{eqnarray}
where $A$ and $B$ are determined by initial conditions for the position $x_c(0)$ and velocity $\dot x_c(0)$ of the particle which  can be deterministic or random. The remaining parameters are defined as
\begin{eqnarray}
	  \Omega^2=\omega_0^2-\gamma_0^2, \quad \omega_0^2=\frac{k}{m}, \quad \gamma_0=\frac{\gamma}{2m}.
\label{scaled}
\end{eqnarray}
The second part $x_d(t)$,   related to the deterministic force $F(t)$,   can be  expressed by  the Green function $G(t, u)$ of equation (1) and takes the form \cite{byron}
\begin{eqnarray}
	 x_d(t) = \int_{0}^{\infty} G(t, u)  F(u) du,
	 \label{xd}
\end{eqnarray}
where the Green function reads  \cite{byron,lam}
\begin{eqnarray}
	 G(t, u) = \Theta(t-u) \, \frac{1}{m \Omega}\, \mbox{e}^{-\gamma_0 (t-u)} \sin [\Omega (t-u)]
	 \label{Green}
\end{eqnarray}
and $\Theta(t)$ stands for the Heaviside step function: $\Theta(t)=1$ for $t>0$ and  $\Theta(t)=0$ for $t<0$. The method of obtaining the Green's function is explained in details in ~\ref{apA}. The form of the Green function (\ref{Green}) is convenient when $\omega_0 > \gamma_0$, i.e. when the system is weakly damped. In the opposite strongly damped regime, when $\gamma_0 > \omega_0$,  one can use the equivalent form with the hyperbolic sine function \cite{lam},
\begin{equation}
	 G(t, u) = \Theta(t-u) \, \frac{1}{m \Omega_0}\, \mbox{e}^{-\gamma_0 (t-u)} \sinh [\Omega_0 (t-u)], \quad \Omega^2_0=\gamma_0^2 - \omega_0^2 =-\Omega^2.  
	 \label{Green1}
\end{equation}
Below, in all intermediate calculations, we will use only one form, namely equation (\ref{Green}).

The third term $x_{\xi}(t)$ is non-deterministic and is related to thermal equilibrium  noise $\xi(t)$. It can also be  expressed by  the Green function $G(t, s)$, namely,
\begin{eqnarray}
	 x_{\xi}(t) =  \sqrt{2\gamma k_B T}\,\int_{0}^{\infty} G(t, u) \xi(u) du.
	 \label{x3}
\end{eqnarray}
Because $\xi(t)$ is a Gaussian process, $x_{\xi}(t)$ as a linear functional of $\xi(t)$ is also Gaussian and in consequence  $x(t)$ is a Gaussian stochastic process. It means that two statistical moments $\langle x(t) \rangle$  and $\langle x^2(t) \rangle$ are sufficient to determine the probability density
$P(x, t)$ of the particle coordinate.

From equation (\ref{xd}) we get
\begin{eqnarray}
\label{xd2}
x_d(t) = \frac{F_0}{m\Omega}\int_0^{t}e^{-\gamma_0(t-u)}\sin[\Omega (t-u)]
\cos(\omega u)~du \nonumber\\
  = \frac{F_0}{m\sqrt{A(\omega)}}\cos (\omega t -\delta)-
  \frac{F_0\omega_0}{m\Omega\sqrt{A(\omega)}}\cos (\omega t -\zeta) 
  \e^{-\gamma_0t},
\end{eqnarray}
where
\begin{eqnarray}
\label{delta}
\delta=\tan^{-1}\frac{2\omega\gamma_0}{\omega^2_0-\omega^2}; \quad \zeta=\tan^{-1}\frac{\gamma_0(\omega_0^2+\omega^2)}{\Omega(\omega^2_0-\omega^2)}; 
\nonumber\\
\label{A}
\quad A(\omega) = (\omega^2 -\omega^2_0)^2 + 4 \gamma_0^2  \omega^2. 
\end{eqnarray}
Now, we can calculate the average value of the particle position $\langle x(t) \rangle$ at time $t$. Because   $\langle \xi(t) \rangle=0$, from Eq.(\ref{x3}) it follows that $\langle x_{\xi}(t) \rangle=0$. Therefore
$\langle x(t) \rangle= \langle x_c(t) \rangle + x_d(t)$. For any initial conditions, $\langle x_c(t) \rangle \to 0$ when $t \to \infty$ and in the stationary state, when all decaying terms vanish, we obtain
\begin{eqnarray}
\label{avxdst}
\langle x(t) \rangle_{st} \equiv X_d(t) =\frac{F_0}{m\sqrt{A(\omega)}}\cos (\omega t -\delta) \nonumber\\ 
= \frac{F_0}{ m A(\omega)}\Big[2\gamma_0  \omega \sin (\omega t)  + (\omega_0^2 -\omega^2) \cos (\omega t) \Big],
\end{eqnarray}
where $X_d(t)$ is the long time limit of the deterministic solution (\ref{xd2}), i.e. a purely periodic, non-decaying part of  (\ref{xd2}).
Let us note that according to the recipe (\ref{averY}),  the period-averaged position in the stationary state is zero, $\overline x=0$. 

\begin{figure}[t]
\centering
\includegraphics[width=0.6\textwidth ]{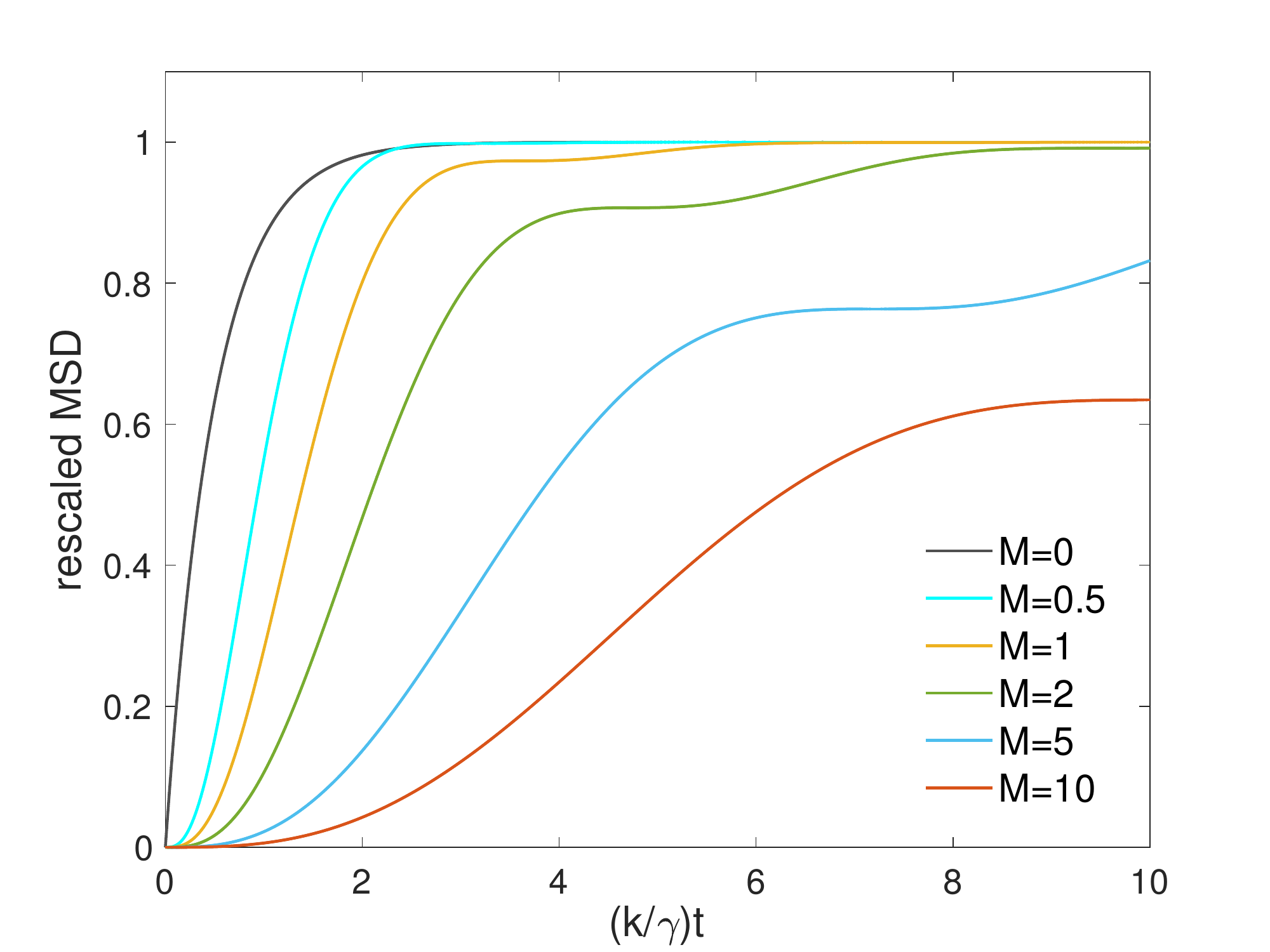}
\caption{ The rescaled mean squared deviation (MSD) 
$\langle \Delta x^2 (t) \rangle/(k_B T/k)$ of the coordinate degree of freedom $x(t)$  versus rescaled time $(k/\gamma)t$ for selected values of the particle rescaled mass $M=(k/\gamma^2) m$. The particles mass only affects the relaxation to the mass-independent steady state  value.}
\label{msdvstime}
\end{figure}

For long times,  the second moment of the position takes the form  
\begin{eqnarray}
\label{x2T}
\langle x^2(t)\rangle &=& \langle [x_c(t) + x_d(t) + x_{\xi}(t)]^2\rangle \nonumber\\
&=&\int_{0}^t \int_{0}^t G(t,u) G(t,u') \big[ F(u)F(u') +2\gamma k_BT \langle \xi(u)\xi(u') \rangle  \big] du du'  
\nonumber \\
&=& X_d^2(t) + 2\gamma k_BT\int_{0}^t  G^2(t,u) du, 
\end{eqnarray}
where we exploited equation (2) for statistical moments of thermal noise $\xi(t)$ and neglected all exponentially decaying terms. 
The integral in equation (\ref{x2T}) can be calculated and in long-time limit,  we again can neglect the expenentially decaying terms to get the relation in the stationary state, 
\begin{equation}
\label{x2(t)}
\langle x^2(t) \rangle_{st}  = \frac{k_B T}{m \omega_0^2} + X_d^2(t). 
\end{equation}
It can be  verified that the mean squared deviation (MSD) of the particle position  is  $\langle (\Delta x)^2\rangle=\langle x_{\xi}^2\rangle$. In Appendix B we present  details of this calculation. The result turns out to be
\begin{equation}
\label{msdx}
\langle \Delta x^2 (t) \rangle = \frac{k_B T }{k}+\frac{ k_B T }{k \Omega^2} \big[ \gamma_0^2 \cos 2 \Omega t - \gamma_0 \Omega \sin 2 \Omega t - \omega_0^2 \big] \e^{-2 \gamma_0 t}. 
\end{equation}
Neglecting the exponentially decaying terms,  we get the following relation in the stationary state,
\begin{equation}
\label{dx2}
\langle \Delta x^2 (t) \rangle_{st}= \frac{k_B T}{k}, 
\end{equation}
which is notably mass independent! In fact the particle's mass only determines the relaxation to the steady state value $k_B T/k$. In Fig.  \ref{msdvstime} we depict the temporal evolution of MSD for several values of the particle mass. As one can see, the larger the particle's mass, the slower the relaxation towards steady state. 
The role of the mass will also be discussed later when we consider the overdamped regime.

\subsection{Velocity of the Brownian particle}

We now come to the particles's velocity analysis. From equation (\ref{x(t)}) it follows that the particle velocity $v(t)= \dot x(t)$ can be decomposed as
\begin{eqnarray}
v(t)=  v_c(t)  + v_d(t)  + v_{\xi}(t).
\label{v(t)}
\end{eqnarray}
The first term can be calculated from equation (\ref{x0}) and for any initial conditions its  average value $\langle v_c(t)\rangle = \langle \dot x_c(t)\rangle \to 0$ if  $t\to\infty$. The second term $v_d(t)$ can be calculated form equation (\ref{xd2}) and in the long time regime it takes the form
\begin{equation}
\label{vd}
 V_d(t) = \lim_{\gamma_0 t \gg 1} v_d(t) = -\frac{F_0 \omega}{m\sqrt{A(\omega)}}\sin (\omega t -\delta).
\end{equation}
The last term is given by:
\begin{eqnarray}
v_{\xi}(t) = \dot x_{\xi}(t) = \sqrt{2\gamma k_B T}\,\int_{0}^{\infty} \frac{\partial G(t,u)}{\partial t} \xi(u) du.
\end{eqnarray}
It can be calculated from Eqs. (\ref{x3}) and (\ref{Green}):
\begin{eqnarray}
	 v_{\xi}(t) = \dot x_{\xi}(t) = -\gamma_0 x_{\xi}(t) \nonumber\\
	 +(1/m)   \sqrt{2\gamma k_B T}\,\int_{0}^{t} e^{-\gamma_0(t-u)}\cos[\Omega (t-u)]  \xi(u) du.
	 \label{v3}
\end{eqnarray}
Its mean value is zero, $\langle v_{\xi}(t) \rangle =0$,  because   both $\langle \xi(t) \rangle=0$ and $\langle x_{\xi}(t) \rangle=0$. As a result, the mean velocity in the steady-state is
\begin{equation}
 \label{Ev}
\langle v(t) \rangle_{st}= V_d(t) =\dot X_d(t).
\end{equation}
One can note that $\langle v(t) \rangle_{st}$ is simply the time derivative of $\langle x(t) \rangle_{st}$ given by equation (\ref{avxdst}).

The second moment of the velocity can be obtained by several methods. One of them is presented in  Appendix B. In the steady-state it reads
\begin{equation}
\label{v2(t)}
\langle v^2(t) \rangle_{st}  = \frac{k_B T}{m } + V_d^2(t).  
\end{equation}
Finally, the mean square deviation of the velocity in the steady-state is 
\begin{equation}
\label{Deltav2(t)}
\langle \Delta v^2(t) \rangle_{st}  = \frac{k_B T}{m }, 
\end{equation}
which is remarkably independent of the spring stiffness $k$! More this value is the one that would be expected in the equilibrium situation (i.e. when $F_0=0$), due to the energy equipartition theorem~\cite{EquPart}.\bigskip

From Eqs. (\ref{x2(t)}) and (\ref{v2(t)}) it follows that in the stationary state the second moment for both position and velocity separates into two additive parts:
\begin{itemize}
\item[(i)] the first part $X_d^2(t)$ for the position (or $V_d^2(t)$ for the velocity) is generated by the deterministic driving force $F(t)$,
\item[(ii)] and the second part $k_B T/k$ for the position (or $ k_B T/m $  for the velocity) is generated by the thermal noise (and is equal to the thermal equilibrium value).
\end{itemize}
This separation takes place because the system is linear and  noise $\xi(t)$ is additive.

\section{Average energy in a stationary state}

In this section, we compute the kinetic and potential energies of the Brownian harmonic oscillator.
The kinetic energy of the oscillator:
\begin{eqnarray}
\label{kinet}
E_k(t) =  \frac{1}{2} mv^2(t)
\end{eqnarray}
is a stochastic quantity.  Its potential energy:
\begin{eqnarray}
\label{poten}
E_p(t) = U(x(t)) = \frac{1}{2} k x^2(t)
\end{eqnarray}
is also a stochastic quantity.  Naturally, the total energy $E(t)$ of the  oscillator is the sum $E(t)=E_k(t)+E_p(t)$. Its average value is determined by the second moments of the coordinate and velocity of the oscillator.
Using the relations (\ref{v2(t)}) and (\ref{vd}), the steady-state average kinetic energy can be expressed by the relation
\begin{eqnarray}
\label{Ek}
\langle E_k(t) \rangle_{st}
=\frac{k_B T}{2} + \frac{F^2_0\omega^2}{2mA(\omega)}\sin^2(\omega t-\delta)
\nonumber \\
=  \frac{k_B T}{2} + \frac{\omega^2 F_0^2 }{4 m A(\omega)} -
 \frac{\omega^2 F_0^2 }{4 m A^2(\omega)} \Bigg[ B(\omega)  \cos(2 \omega t) - C(\omega) \sin(2 \omega t)  \Bigg],  
\end{eqnarray}
where  $A(\omega)$ is defined in equation (\ref{A}) and 
\begin{eqnarray}
\label{R-}
B(\omega) = ( \omega^2 - \omega_0^2 )^2 - 4 \gamma_0^2 \omega^2, \qquad 
 C(\omega) = 4 \omega \gamma_0 ( \omega^2 - \omega_0^2 ). 
\end{eqnarray}
From Eqs. (\ref{x2(t)}) and (\ref{avxdst}) it follows that the steady-state average potential energy has the form
\begin{eqnarray}
\label{Ep}
\langle E_p(t) \rangle_{st} 
=  \frac{k_B T}{2} +  \frac{F^2_0\omega_0^2}{2mA(\omega)}\cos^2(\omega t-\delta)
\nonumber\\
= \frac{k_B T}{2} + \frac{\omega_0^2 F_0^2 }{4 m A(\omega)} +  
\frac{ \omega_0^2 F_0^2 }{4m A^2(\omega)} \Bigg[B(\omega) \cos(2 \omega t) - C(\omega) \sin (2 \omega t) \Bigg]. 
\end{eqnarray}
The total averaged energy of the oscillator in the stationary state is
\begin{eqnarray}
\label{Ec}
\langle E(t) \rangle_{st} &=&  k_B T + \frac{( \omega^2 + \omega_0^2)F_0^2 }{4 m A(\omega)} \nonumber\\ 
&+& \frac{(\omega_0^2 - \omega^2)F_0^2  }{4m A^2(\omega)} \Bigg[ B(\omega) \cos(2 \omega t) - C(\omega) \sin (2 \omega t) \Bigg].   
\end{eqnarray}
Both kinetic energy and potential energy consist of two terms: thermal equilibrium  energy $k_BT/2$ and the driving-generated part proportional to its amplitude squared  $F_0^2$. In turn, the driving-generated part consists of two terms: the part constant in time  and the time-periodic part. For the latter, its  mean value over the driving $\mathsf{T}$-period according to the rule (\ref{averY}) is zero. In contrast to the position $\langle x(t) \rangle_{st}$ and velocity $\langle v(t) \rangle_{st}$, which are $\mathsf{T}$-periodic function of time, the energy  is periodic with respect to time but now the period is $\mathsf{T}/2 =\pi/\omega$.

\begin{figure}[t]
	\centering
\includegraphics[width=0.49\linewidth]{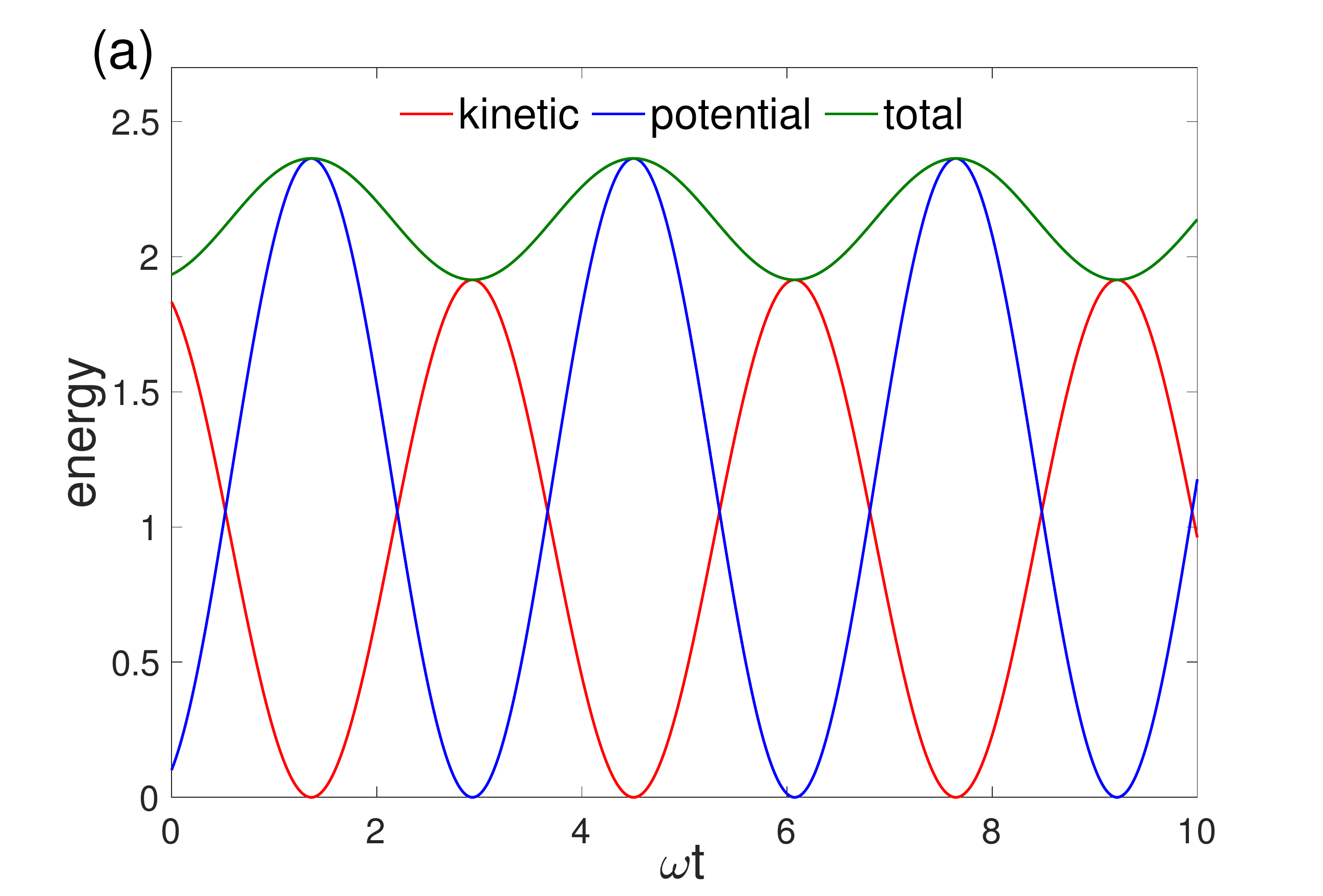}
\includegraphics[width=0.49\linewidth]{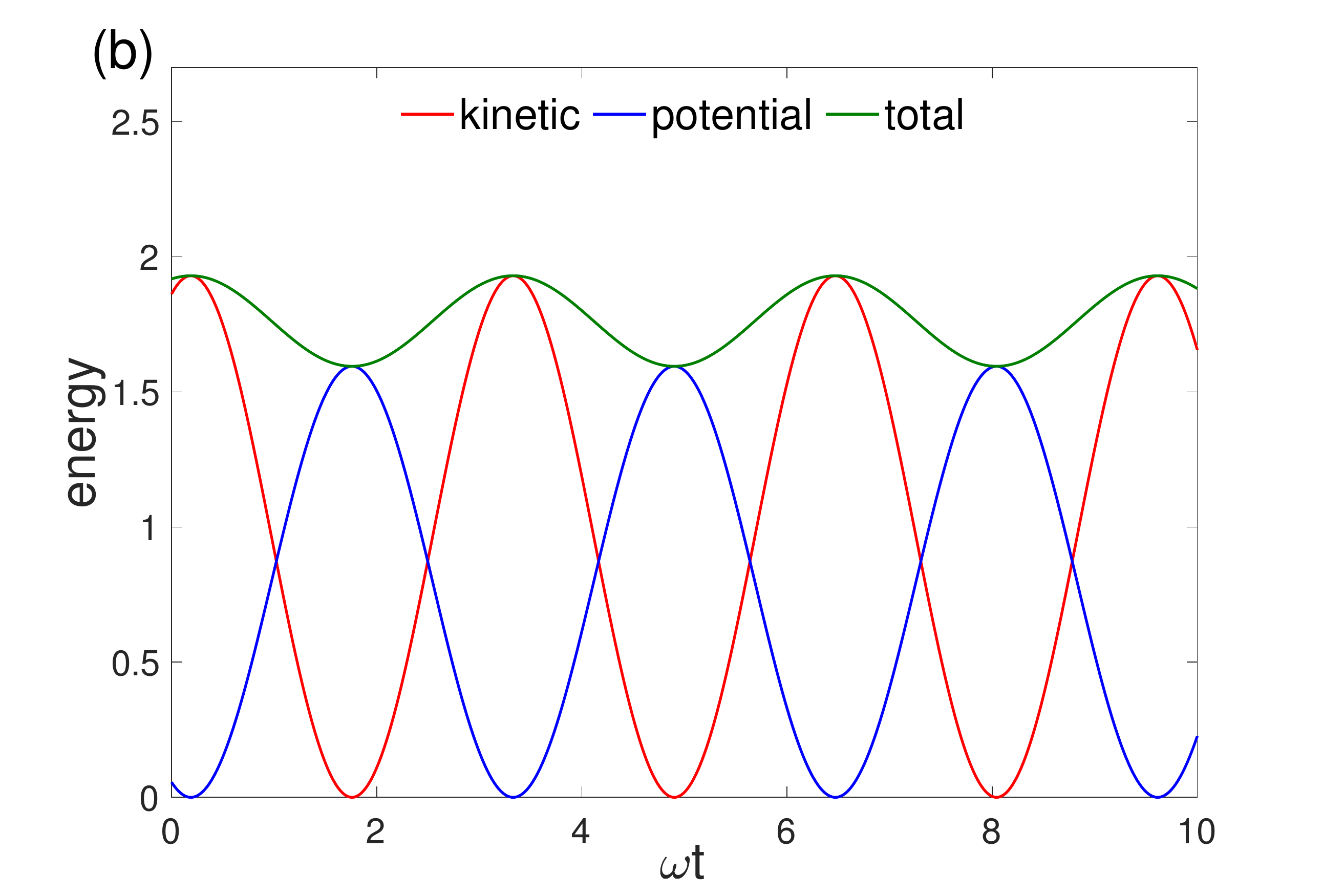}
\includegraphics[width=0.49\linewidth]{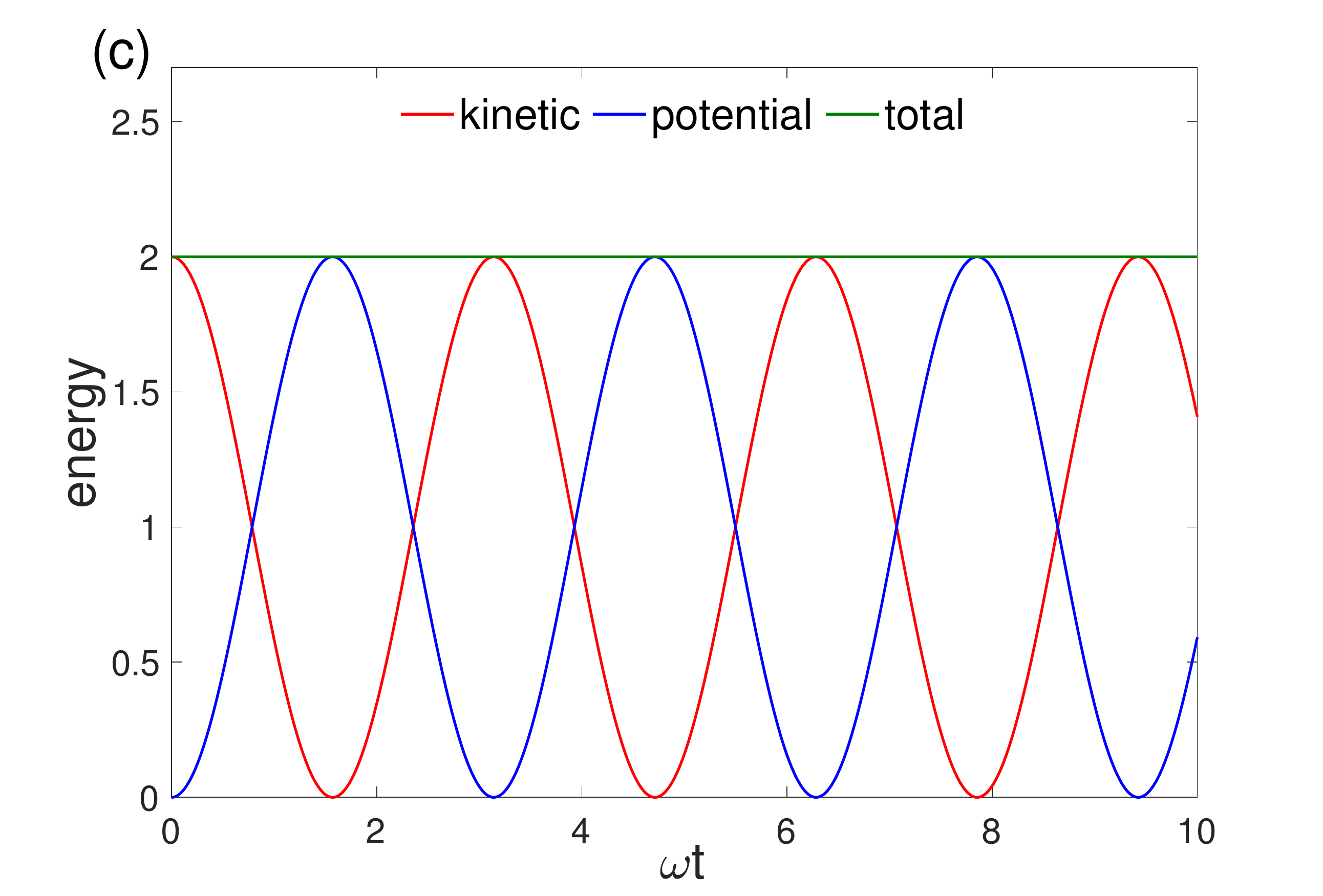}
\caption{ The driving-generated part of the dimensionless averaged kinetic, potential and total energy in the stationary regime. The dimensional energy is divided by the characteristic energy
$E_0 = F_0^2/4m\omega_0^2=F_0^2/4k$. The dimensionless time is $t'=\omega_0 t$.  The rescaled friction coefficient is $\Gamma=\gamma/\sqrt{mk}=1$.  Panel a: $\omega / \omega_0=0.9$. The maximal value of the potential energy is greater than the maximal value of the kinetic energy.  Panel b: $\omega / \omega_0=1.1$. The maximal value of the potential energy is smaller than the maximal value of the kinetic energy. Panel c: $\omega / \omega_0=1$. The maximal value of the potential energy is equal to the maximal value of the kinetic energy and the total energy is conserved in time. }
	\label{fig2}
\end{figure}

Time-changes of energy in the stationary state  are depicted in Fig.~\ref{fig2}. In all panels of  this figure, the rescaled kinetic energy is: $\varepsilon_k(t) = [\langle E_k(t) \rangle_{st} -k_BT/2]/E_0$;
the rescaled potential  energy is: $\varepsilon_p(t) = [\langle E_p(t) \rangle_{st} -k_BT/2]/E_0$;
and the rescaled total energy is: $\varepsilon(t)= \varepsilon_k(t)+\varepsilon_p(t)$, where the characteristic energy  	$E_0 = F_0^2/4m\omega_0^2=F_0^2/4k$.
One can note that the minimal values of the kinetic and potential energies are the equilibrium values $k_BT/2$. However, their maximal values depend on the ratio $\omega/\omega_0$ of two characteristic frequencies, namely, the frequency $\omega$ of the time-periodic force $F(t)$ and the eigenfrequency $\omega_0$ of the oscillator.  If $\omega < \omega_0$ (panel a in Fig.~\ref{fig2})  then the amplitude of the potential energy is greater than the amplitude of the kinetic energy. In turn, if $\omega > \omega_0$ (panel b in Fig.~\ref{fig2}) then the amplitude of the kinetic energy is greater than the amplitude of the potential energy. In both cases, the total energy oscillates in time. Finally, if $\omega = \omega_0$ (panel c in Fig.~\ref{fig2}), the total energy does not depend on time, it is constant! Remember that the system is open, not isolated. There is dissipation of energy due to contact with thermostat of temperature $T$ and there is a pumping of energy by the time-periodic force $F(t)$.  Under some specific conditions (here $\omega = \omega_0$), these two processes can lead to the {\it conservation of total energy in the noisy system}.

\section{Energy balance}

We now derive equations for the energy balance. To do this, we first rewrite  equation (\ref{Lan}) in the Ito-type form of stochastic differential equations:
\begin{eqnarray}
\label{dx}
dx &=& v dt\;,\\
m \, dv &=&  [- \gamma v - kx + F(t)] dt + \sqrt{2 \gamma k_BT} dW(t)\;,
\label{dv}
\end{eqnarray}
where  formally  $dW(t)=\xi(t) dt$ and $W(t)$ is the Wiener process of Gaussian statistics with zero mean
$ \langle dW(t) \rangle = 0$ and the second moment $\langle [dW(t)]^2 \rangle = dt$. Next, we employ the Ito-differential calculus to the functions $E_p(x)$ and $E_k(v)$, namely,
\begin{eqnarray}
\label{dEp}
dE_p =\frac{dE_p}{dx} dx +  ... =  kxvdt,   \\
dE_k = \frac{dE_k}{dv} dv + \frac{1}{2} \frac{d^2E_k}{dv^2} dv dv  + ... \nonumber\\
 =
\left(- \gamma v^2 -kx v + v F(t) \right) dt
+ \sqrt{2 \gamma k_BT} v~dW(t) + \frac{\gamma k_BT}{m}~[dW(t)]^2, \nonumber \\
\label{dEk}
\end{eqnarray}
where we have only kept the terms that at most are first order in $dt$. We then perform the ensemble averaging $\langle ... \rangle$ over all realizations of the Wiener process $W(t)$ and over all initial conditions $\{x(0), v(0)\} $. As a result we obtain the rate of change of the potential and kinetic energies in the form:
\begin{eqnarray}
\label{rate}
\frac{d}{dt} \langle E_p (t)\rangle &=& k \langle x(t) v(t)  \rangle ,   \\
\frac{d}{dt} \langle E_k(t) \rangle &=& -\gamma  \langle v^2(t) \rangle -
k \langle x(t) v(t)  \rangle +  \langle v(t) F(t)  \rangle  + \frac{\gamma k_BT}{m},
\end{eqnarray}
where for the part containing the Wiener process  in equation (\ref{dEk}) we used  the Ito-martingale property, i.e. $\langle v(t) dW(t)\rangle =0$. By adding these two equations together, we get an equation for the energy balance. It reads:
\begin{eqnarray}
\label{E}
\frac{d}{dt} \langle E(t) \rangle = -\gamma  \langle v^2(t) \rangle
+  F(t) \langle v(t) \rangle  + \frac{\gamma k_BT}{m}.
\end{eqnarray}
We conclude that there are three processes responsible for the energy change:
(i) the first term in the r.h.s. of equation (\ref{E}) is always negative and  represents the rate of energy loss due to dissipation (friction of the fluid); (ii) the second term describes pumping of energy to the system by the external force $F(t)$ and (iii)  the third term is always positive and  characterizes the energy supplied to the system by thermal equilibrium fluctuations. We recall that in thermal equilibrium (i.e. when $F_0=0$), the theorem on the equipartition of energy states that $\langle E_k \rangle_{\it eq} =k_BT/2$, from which it follows that $\langle v^2 \rangle_{\it eq} = k_BT/m$. Therefore equation (\ref{E}) can be rewritten in the form:
\begin{eqnarray}
\label{EE}
\frac{d}{dt} \langle E(t)\rangle = -\gamma [ \langle v^2(t) \rangle  -
\langle v^2 \rangle_{\it eq}] +  F(t) \langle v(t) \rangle.
\end{eqnarray}
Let us note that when $F(t)=0$, in the stationary state (which is thermal equilibrium), the l.h.s. tends to zero and  we retrive $\langle v^2(t) \rangle \to \langle v^2 \rangle_{\it eq} = k_BT/m$ for any initial conditions.\\
Now, we perform the time averaging of equation (\ref{EE}) according to the prescription (\ref{averY}). Let us remember that in the stationary state the average values of any functions of the particle coordinate $x(t)$ and its  velocity $v(t)$ are periodic functions of time and therefore  the left hand side equation (\ref{EE}) vanishes after time averaging. Indeed, by definition:
\begin{eqnarray} \label{left}
\int_t^{t+2\pi/\omega} \frac{d}{du} \langle E(u) \rangle_{st} \, du &=& \frac{m}{2} [\langle v^2(t+2\pi/\omega)\rangle_{st}  - \langle v^2(t)\rangle_{st}  ] \nonumber\\
&+&  [\langle U(x(t+2\pi/\omega))\rangle_{st}  -  \langle U(x(t))\rangle_{st}  ] = 0. 
\end{eqnarray}
As a consequence, in the stationary regime the mean power $\overline{\mathcal P_{in}}$ delivered to the system by the external force $F(t)$ over the period $T=2\pi/\omega$ is expressed by the relation
\begin{eqnarray}
\overline{\mathcal P_{in}} = \overline{F(t)\langle v(t) \rangle} = \gamma  \left[ \overline{\langle v^2(t) \rangle}   - \langle v^2 \rangle_{eq}\right] =
\frac{2}{\tau_1}  \left[\overline{\langle E_k \rangle} - \langle E_k \rangle_{eq}\right],
\label{e2}
\end{eqnarray}
where the characteristic time $\tau_1 = m/\gamma = 1/(2\gamma_0)$ is the relaxation time of the particle velocity for the system when $F(t)=0$. This relation has an appealing physical interpretation: the input power from the external force $F(t)$ to the system determines the difference between the averaged non-equilibrium  $\langle E_k \rangle$ and equilibrium $\langle E_k \rangle_{\it eq}$ values of the kinetic energy of the Brownian particle. Moreover, the input power depends not only on the force itself (i.e. on the amplitude $F_0$ and frequency $\omega$) but also on properties and parameters of the system:  the friction coefficient $\gamma$ and the mass  $m$. In contrast, the energy supplied by thermal fluctuations does not depend on the external force  but only on $T, \gamma$ and $m$.

\begin{figure}[t]
\centering
\includegraphics[width=0.49\linewidth]{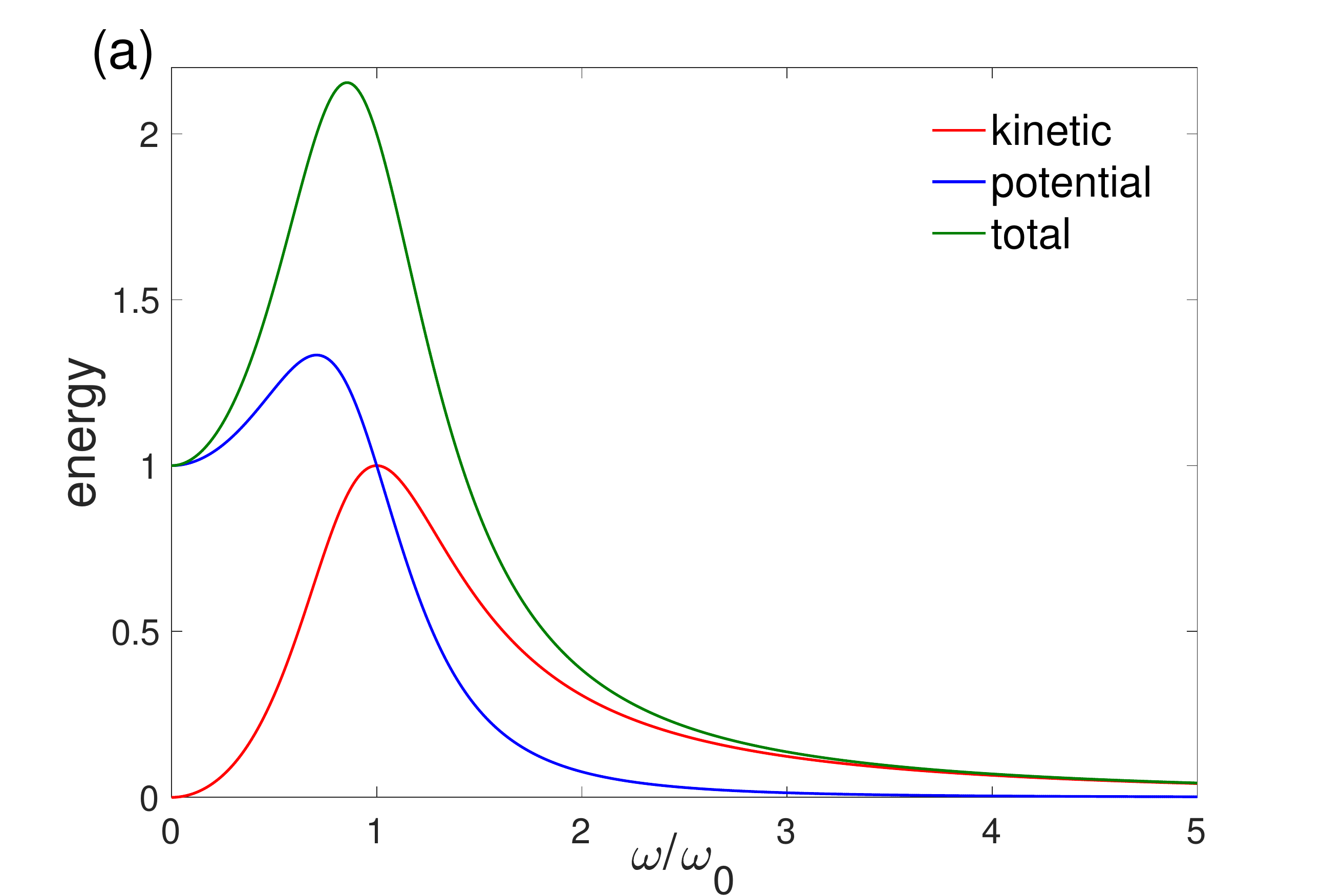}
\includegraphics[width=0.49\linewidth]{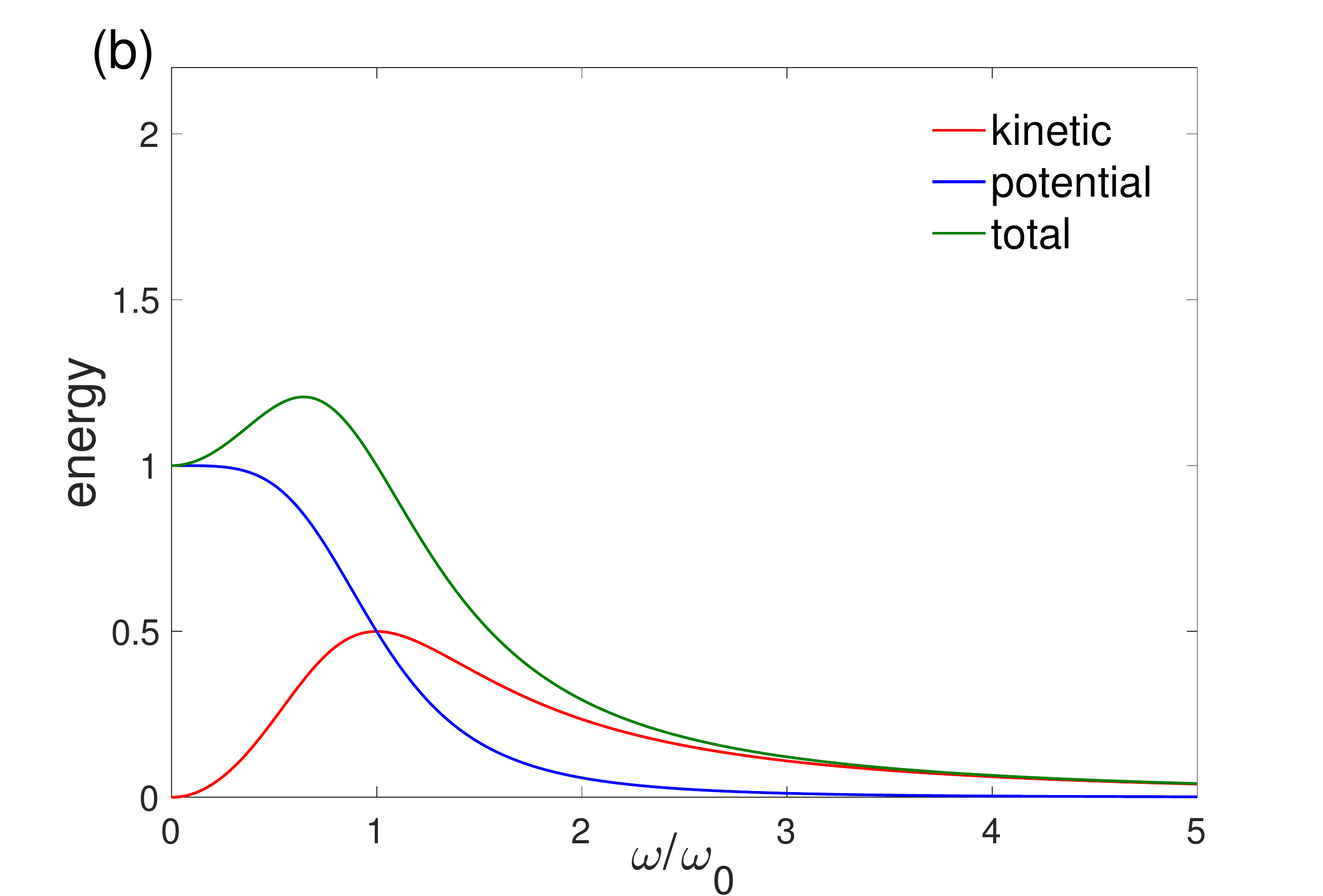}
\includegraphics[width=0.49\linewidth]{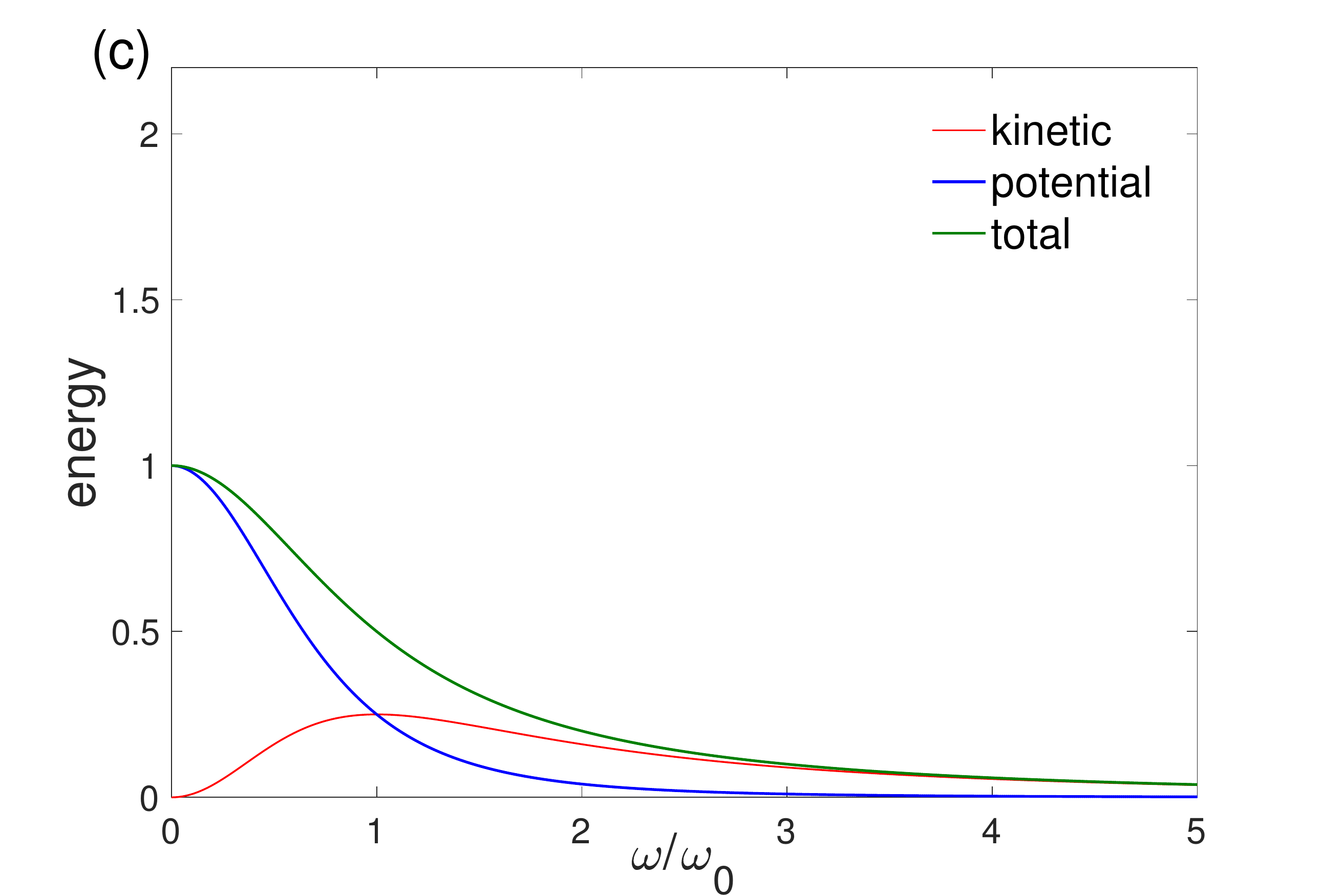}
\caption{  The driving-generated part of the rescaled energies as a function of the rescaled frequency $\Omega= \omega/\omega_0$ of the external time-periodic force. Kinetic, potential and total energy in stationary regime  averaged over the period of the external force, cf. Eqs. (\ref{ek})-(\ref{ec}). The rescaled friction coefficient
$\Gamma = \gamma/\sqrt{m k} = 1$ (panel a);  $\Gamma = \sqrt{2}$ (panel b) and $\Gamma = 2$ (panel c). }
\label{fig3}
\end{figure}

\section{Period-averaged energy}

In the non-equilibrium stationary state all averaged forms of energy are periodic functions of time. Time independent characteristics can be obtained from the formulas (\ref{Ek})-(\ref{Ec}) by  additional averaging  over the  period  $\mathsf{T} =  2\pi / \omega$ of  the time-periodic force $F_0 \cos(\omega t)$. The averaging scheme is  according to the rule (\ref{averY}). In this way we obtain the period-averaged kinetic energy in the form
\begin{equation} \label{ek}
\overline{ E_k }= \frac{k_B T}{2} +
\frac{F_0^2}{4m} \;
\frac{ \omega^2 }{(\omega^2 -\omega_0^2)^2 + (\gamma  \omega/m)^2}.
\end{equation}
The period-averaged potential energy reads
\begin{equation} \label{ep}
\overline{ E_p } = \frac{k_B T}{2} + \frac{F_0^2}{4m} \;
\frac{ \omega_0^2 }{(\omega^2 -\omega_0^2)^2 + (\gamma  \omega/m)^2}
\end{equation}
and the total period-averaged energy is given by the expression
\begin{equation} \label{ec}
\overline{ E } =  k_B T + \frac{F_0^2}{4m} \;
\frac{ \omega^2 + \omega_0^2}{(\omega^2 -\omega_0^2)^2 + (\gamma  \omega/m)^2}.
\end{equation}
The period-averaged input power can be obtained form  equation (\ref{e2}) and the result is
\begin{equation}
 \label{Pin}
\overline{\mathcal P_{in}} = \frac{\gamma F_0^2}{2m^2} \;
\frac{ \omega^2 }{(\omega^2 -\omega_0^2)^2 + (\gamma  \omega/m)^2}. 
\end{equation}
We recall that the time averaged input power $\overline{\mathcal P_{in}}$ is independent of temperature $T$ of the system. It is not the case for non-linear systems \cite{NJPspie} for which the input power is temperature dependent. 

\begin{figure}[t]
\centering
\includegraphics[width=0.49\linewidth]{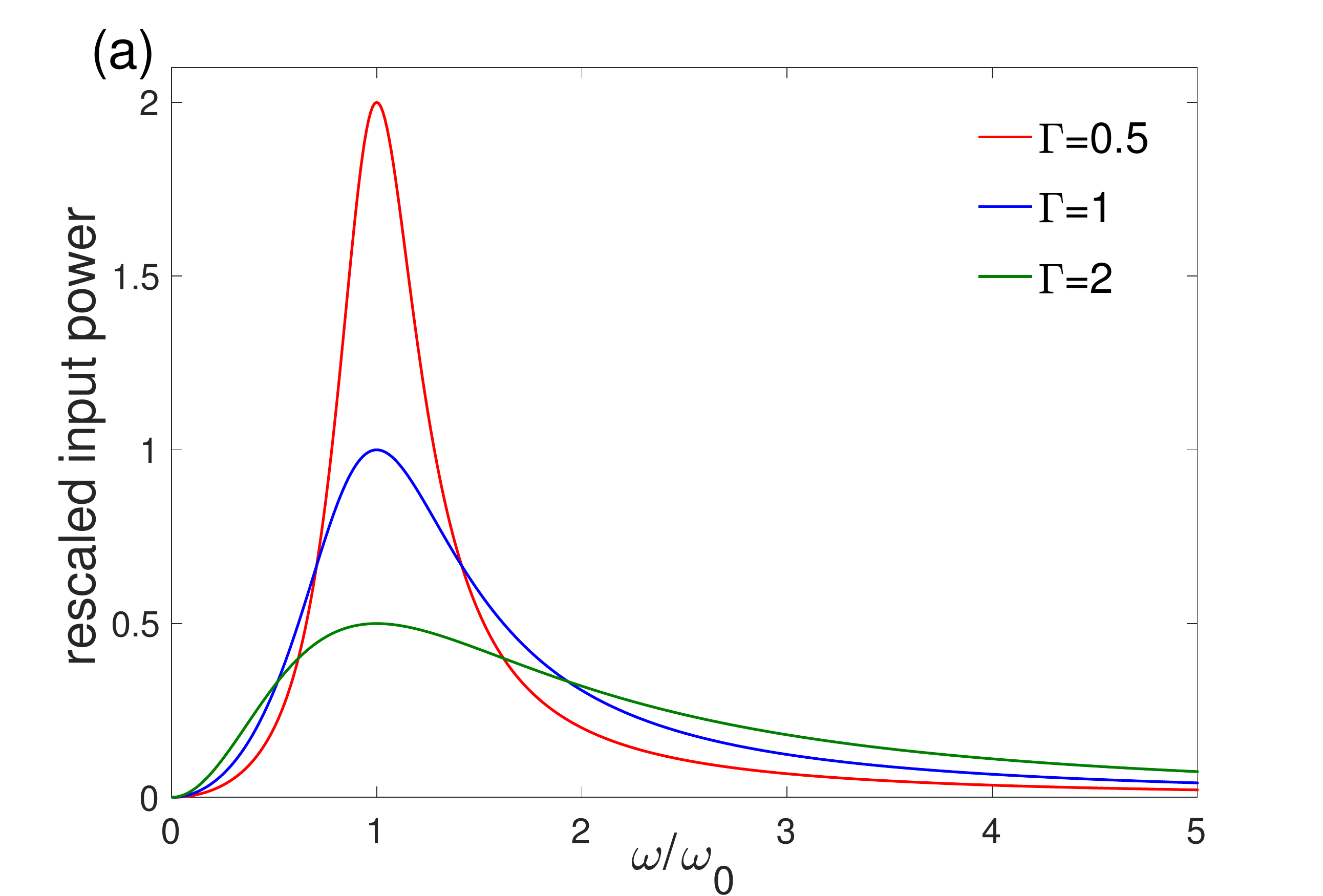}
\includegraphics[width=0.49\linewidth]{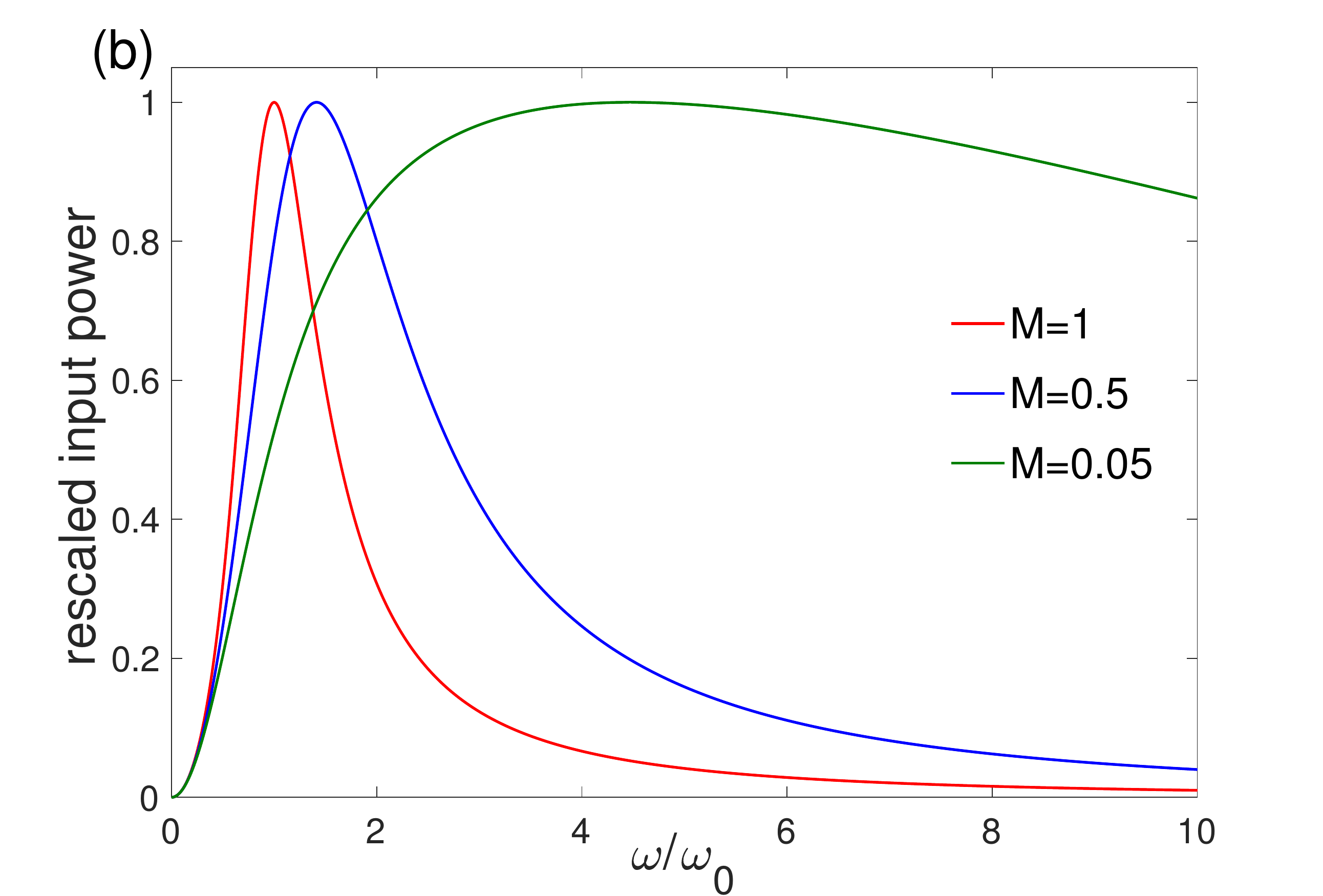}
\caption{ The rescaled input power as a function of the rescaled frequency $\Lambda$ of the external time-periodic force. Panel a: for selected values of the rescaled friction coefficient $\Gamma = \gamma/\sqrt{m k} =0.5, 1, 2$ and the rescaled frequency is $\Lambda=\omega/\omega_0$. Panel b: for selected values of the rescaled mass $M=mk/\gamma^2 =1, 0.5, 0.05$ and the scaled frequency is $\Lambda=(\gamma/k) \omega$. }
\label{fig4}
\end{figure}

Now, we analyse properties of these time-independent mean energies.
The dependence of energy on the rescaled frequency $\Lambda =\omega/\omega_0$ of the external force is shown in Fig.~\ref{fig3}. We observe that:
\begin{itemize}
\item[\textbf{A.}] The kinetic energy is zero at  zero driving frequency $\Lambda=0$ and approaches zero when $\Lambda \to \infty$. It means that there must be  an optimal value of the driving force frequency  for which the kinetic energy is maximal. Calculation of derivative shows that the maximal value is for the resonance frequency  $\Lambda=1$.
\item[\textbf{B.}] The shape of the potential energy depends on the rescaled friction coefficient $\Gamma=\gamma/\sqrt{m k}$: If $\Gamma^2 \ge 2$, the potential energy takes a bell-shape form of  the maximal value at $\Lambda =0$. However, if $\Gamma^2 < 2$, there is one  minimum at $\Lambda =0$ and the maximum  at $\Lambda^2 = 1-\Gamma^2/2 $.
\item[\textbf{C.}] The shape of the total energy also depends on the rescaled friction coefficient: If $\Gamma^2 \ge 3$, the bell-shaped total energy is maximal  at $\Lambda =0$ and tends to zero when the frequency increases.   In turn, if $\Gamma^2 < 3$, it exhibits one  minimum at $\Lambda =0$ and the   maximal value  at $\Lambda ^2 = \sqrt{4-\Gamma^2} -1$.
\end{itemize}
The dependence of the input power on the rescaled friction coefficient and the rescaled mass is shown in Fig.~\ref{fig4}.  In panel a, the scaled frequency is $\Lambda=\omega/\omega_0$ and the rescaled input power is given by the formula:
\begin{equation}
 \label{Ping}
\frac{\overline{\mathcal P_{in}}}{\mathcal P_{0}} =
\frac{ \Gamma \Lambda^2 }{(\Lambda^2 - 1)^2 + \Gamma^2 \Lambda^2},
\end{equation}
where the characteristic power is $\mathcal P_{0}=F_0^2/(2\sqrt{mk})$. The minimum input power is always at $\Lambda=0$,  the maximum is always at the resonance $\Lambda =1$ and does not depend on $\Gamma$. In panel b, the scaled frequency is $\Lambda=(\gamma/k) \omega$  and the rescaled input power now reads:
\begin{equation}
 \label{PinM}
\frac{\overline{\mathcal P_{in}}}{\mathcal P_{1}} =
\frac{  \Lambda^2 }{(M \Lambda^2 - 1)^2 +  \Lambda^2},
\end{equation}
where the characteristic power is $\mathcal P_{1}=F_0^2/(2\gamma)$ and the rescaled mass is  $M=mk/\gamma^2 $. The minimum input power is always at $\Lambda=0$. The maximal input power is $\mathcal P_{1}$ which  is located at $\Lambda= \pm 1/\sqrt{M}$.

\section{The Jarzynski-type work relation }

Many physical observables undergo random fluctuations. The coordinate, velocity, kinetic and potential energies of a Brownian particle are generic examples of such observables. Work and fluctuation relations constitute an additional  basic characterization of statistical properties of the system. In this section, we present the counterpart of the Jarzynski work relation for the considered system. We have to stress that  what we present is not an  exact Jarzynski relation because  conditions for the  Jarzynski theorem to be hold  are not fulfilled and the interpretation is not in terms  of a free energy, the notion  which is defined for systems in thermal equilibrium but not for non-equilibrium ones. 

Although the conception of work seems to be well known,  there is no consensus on its precise and general definition, see discussion in Refs. 
\cite{CHT11,Jarz2,HanTal}. Here we consider the {\it inclusive} work  defined as  
\begin{equation}
\label{inclu1}
W_{\tau_0}(\tau) = -\int_{\tau_0}^{\tau_0+\tau} \dot F(t) x(t) dt = \omega F_0 \int_{\tau_0}^{\tau_0+\tau} \sin(\omega t) x(t) dt, 
\end{equation}
where $x(t)$ is given by equations (5)-(13). It is the work performed by the external force $F(t)$ on the system in the time interval $(\tau_0, \tau_0+\tau)$. We are interested in the limit of long times when the asymptotic  state is reached. Therefore the regime $\tau_0 >>1$ should  be considered. 

The coordinate  $x(t)$, via the relation (\ref{x3}), is a linear functional of the Gaussian process $\xi(t)$. Therefore  for each fixed time  $\tau_0$ the work (\ref{inclu1}) as a function of time $\tau$ 
is a stochastic Gaussian process. However, because of the time-periodic driving $F(t)$,  it is a non-stationary random process and therefore the statistical properties of the work $W_{\tau_0}(\tau)$ can depend on the reference time $\tau_0$! It is a radically different case than for  stationary processes.  \\
The  probability distribution of the work $W_{\tau_0}(\tau)$ is a Gaussian function in the form \cite{papoulis} 
\begin{equation}
\label{Pw}
P_{\tau_0}(w, \tau) = \frac{1}{\sqrt{2\pi \sigma_{\tau_0}^2(\tau)}} \exp[-(w -\mu_{\tau_0}(\tau))^2/ 2\sigma_{\tau_0}^2(\tau)],  
\end{equation}
where 
\begin{equation}
\label{stat01}
\mu_{\tau_0}(\tau) = \langle W_{\tau_0}(\tau)\rangle, \quad 
\sigma_{\tau_0}^2(\tau) =  \langle W_{\tau_0}^2(\tau)\rangle - \langle W_{\tau_0}(\tau)\rangle^2
\end{equation}
are the mean value and variance of the work, respectively.
 
The characteristic function $C_{\tau_0}(z, \tau)$ of the Gaussian stochastic process  $W_{\tau_0}(\tau)$  
is well known, namely \cite{papoulis}, 
\begin{equation}
\label{Cw}
C_{\tau_0}(z, \tau) = \langle \e^{iz W_{\tau_0}(\tau)}\rangle =
 \exp\left[i\mu_{\tau_0}(\tau) z -\frac{1}{2} \sigma_{\tau_0}^2(\tau) z^2\right].  
\end{equation}
Now, let us assume that $z=i\beta$, where $\beta =1/k_B T$. Then it takes the form  
\begin{equation}
\label{Jarz1}
 \langle \e^{-\beta W_{\tau_0}(\tau)}\rangle = \e^{-\beta \Delta F(\tau_0, \tau)}, 
 \qquad \Delta F(\tau_0, \tau)=\mu_{\tau_0}(\tau) - \beta \sigma_{\tau_0}^2(\tau)/2.   
\end{equation}
This characteristic function of the work $W_{\tau_0}(\tau)$  for the imaginary argument $z=i \beta$ is just a counterpart of the Jarzynski relation: 
\begin{equation}
\label{Jarz}
 \langle \e^{-\beta W}\rangle =
 \e^{- \beta \Delta F},    
\end{equation}
where $\Delta F$ is the free energy difference between the final and initial equilibrium states. 
 The argument $\Delta F(\tau_0, \tau)$ of the exponent in r.h.s. of (\ref{Jarz1}) formally corresponds to $\Delta F$ in (\ref{Jarz}) but cannot be interpreted as a free energy of the system. 
 
  In  Appendix C,   we calculate both $\mu_{\tau_0}(\tau)$ and $\sigma^2_{\tau_0}(\tau)$ which allow to calculate $\Delta F(\tau_0, \tau)$. As a result we get 
\begin{eqnarray}
\label{deltaF}
\Delta F(\tau_0, \tau) =  \frac{1}{2m} \frac{\omega^2 -\omega_0^2}{A(\omega)} \left[F^2(\tau_0 +\tau) - F^2(\tau_0)\right] \nonumber\\
+ \frac{1}{2\omega} \overline{\mathcal P_{in}} 
 \left\{\sin\left[\omega(\tau_0+\tau)\right]- \sin[2\omega (\tau_0 +\tau)] - \sin(\omega \tau_0) + \sin(2\omega \tau_0)\right\} \nonumber\\  
 - \frac{1}{2} \overline{\mathcal P_{in}} \tau_1 C_1 \left\{\cos\left[\omega(\tau_0+\tau)\right] -\cos(\omega \tau_0)\right\} 
 -\frac{1}{2} \overline{\mathcal P_{in}} \,\tau_1  \left[ e^{-\gamma_0 \tau} I(\tau_0, \tau) - I(\tau_0, 0)\right],  \nonumber\\
 \ \ \ 
\end{eqnarray}
where $\overline{\mathcal P_{in}}$ is defined by equation (\ref{Pin}) and the remaining notation is explained in Appendix C. 

In the literature \cite{cohen,janavar,velasco},  the relation (\ref{Jarz1}) is called a stationary state fluctuation relation for the total work performed during the time $\tau$. However, in Ref. \cite{cohen,velasco},  a special case of the stationary state is considered, namely, when both statistical moments $\mu_{\tau_0}(\tau)$ and  $\sigma_{\tau_0}^2(\tau)$ do not depend on the reference time $\tau_0$. In our case, both  $\mu_{\tau_0}(\tau)$ and  $\sigma_{\tau_0}^2(\tau)$ do  depend on $\tau_0$, cf. equation (C.1) and (C.7) in Appendix C.  Moreover, the well-known equality  
$\beta \sigma_{\tau_0}^2(\tau) = 2 \mu_{\tau_0}(\tau)$ does not hold, even in the limiting case  $\tau\to \infty$. It is only one term  $ \overline{\mathcal P_{in}} \; \tau$ in $\mu_{\tau_0}(\tau)$  and one term $2 \overline{\mathcal P_{in}} \; \tau$ in $\beta \sigma_{\tau_0}^2(\tau)$ 
which connotes this equality. 

\section{Strong damping regime}

The regime of strong damping  has played  an important role in study of deterministic and noise-assisted dynamical systems  and is usually relevant for experimental micro-systems. The common way to treat the overdamped dynamics is to put mass zero, $m=0$, in Eq.~\ref{Lan}. 
However, this approach is not fully rigorous because of several reasons. First, the mass of the particle is never exactly zero: it can be 'small', but then remains the question 'small' in comparison to what? Second,  the energy  balance cannot be fully evaluated since some  terms do not exist separately but only in specific combinations. It can give rise to squares of the Dirac $\delta$-functions which do not exist \cite{jung}. In other method  the order of magnitude of the dimensionless quantity $m\ddot{x}/\gamma \dot{x}$ is considered. The problem with this criterion is that the fraction is not constant in time and it is even worse because it can diverge to infinity. A more correct procedure  is to identify  regimes  where 'something' is small or   'something' is large. Moreover, one can use a technique of mathematical sequences and try to obtain an exact limit of a considered  sequence. We present the following scheme: We  transform equation (\ref{Lan}) to its dimensionless form and  work in dimensionless variables, because from the physical point of view only relations between scales of
time, length  and energy play a role and not their absolute values.
For the system defined by Eq.~\ref{Lan}, one can identify four characteristic times:
\begin{itemize}
\item[\textbf{1.}]
 The relaxation time of the velocity $\tau_1=m/\gamma $, which can be extracted from the special case of equation (1), namely, $m\dot{v} = -\gamma v$.
 \item[\textbf{2.}]
The relaxation time of the position $\tau_2 =\gamma/k$ deduced from the equation  $\gamma \dot x = -kx$.
\item[\textbf{3.}]
 The reciprocal of the frequency of small oscillations $\tau_3= \sqrt{m/k}$ obtained from the equation $m\ddot x = -kx$.
 \item[\textbf{4.}] The periodicity time related to the frequency of the external driving $\tau_4=1/\omega$.
\end{itemize}

 To analyze strong damping regime, one has to choose the characteristic time $\tau_2$ as a scaling time since it does not contain the particle mass $m$ and therefore the dimensionless time is defined as $s=t/\tau_2$. We also introduce the dimensionless position $y=x/l_0$, where $l_0$ is some characteristic distance. In the case of the harmonic potential, the only reasonable length scale is defined by the equipartition theorem $\langle E_p\rangle=\langle kx^2/2\rangle = k_BT/2 =kl_0^2/2$.  In consequence  a characteristic distance is $l_0= \sqrt{k_BT/k}$. Then the dimensionless Langevin  equation corresponding to Eq.~(\ref{Lan}) takes the form:
\begin{equation}
\label{Newton}
M \frac{d^2y}{ds^2}  + \frac{dy}{ds} + y =  \tilde F(s) + \sqrt{2D_0} \, \tilde \xi(s),
\end{equation}
where the rescaled quantities  read
\begin{equation}
\label{rescal}
M = \tau_1/\tau_2 = mk/\gamma^2, \quad
\tilde F(s) = F(\tau_2 s)/kl_0, \quad D_0=k_BT/kl_0^2 =1
\end{equation}
The rescaled Gaussian noise $\tilde \xi(s)$ has the same statistics as $\xi(t)$ in equation (2). Note that (i): in the left-hand-side of equation (\ref{Newton}) there is only one parameter $M$ which is the same as in equation (\ref{PinM}) and (ii): the intensity $D_0$ of the rescaled thermal noise is 1.  The parameter $M$ is a ratio of two characteristic times of the system  and if $\tau_1 \ll \tau_2$ then $M\ll1$,  and we can neglect the inertial term.
Observe  that the inequality $M \ll 1$ means that $mk/\gamma^2 \ll1$ and it  defines  a real physical regime of a strong damping regime.  It depends not only on mass $m$ and friction coefficient $\gamma$ but also on the stiffnes $k$ of the potential trap! In literature, this fact is often ignored. 

Now, we can rewrite Eqs. (\ref{ek})-(\ref{ep}) in the dimensionless form . The rescaled kinetic energy is (here we consider only the contribution of the forcing on the energies, so we subtract the $k_BT/2$ contribution due to the thermal agitation):
\begin{equation} \label{r_ek}
\varepsilon_k = [\overline{E_k} -k_BT/2]/E_0 =
\frac{  M \Lambda^2 }{(M \Lambda^2 - 1)^2 +  \Lambda^2},
\end{equation}
where the rescaled frequency $\Lambda = (\gamma/k)\omega$ and $E_0=F_0^2/4k$.
For the  potential energy we get
\begin{equation} \label{r_ep}
\varepsilon_p = [\overline{E_p} -k_BT/2]/E_0 =
\frac{ 1 }{(M \Lambda^2 - 1)^2 +  \Lambda^2}.
\end{equation}
The rescaled input power is given by equation (\ref{PinM}). We find that
if $M$ decreases to zero, the rescaled  kinetic energy $\varepsilon_k$ decreases to zero. In this limit, the potential energy $\varepsilon_p$ takes the similar form like the Cauchy probability distribution, namely,
\begin{equation} \label{as_ep}
\varepsilon_p = \frac{ 1 }{1 +  \Lambda^2}.
\end{equation}
 Similar behavior exhibits  the total energy because kinetic energy is zero. As follows from equation (\ref{PinM}), if $M\to 0$ the input power assumes the form
\begin{equation}
 \label{Pin0}
\frac{\overline{\mathcal P_{in}}}{\mathcal P_{1}} =
\frac{  \Lambda^2 }{1+\Lambda^2}.
\end{equation}
It is zero for  zero frequency (i.e. when the force $F(t)$ is constant) and $\overline{\mathcal P_{in}}$ saturates to the value $\mathcal P_{1}$ when the frequency $\Lambda \sim\omega \to \infty$. However, it is not a correct physical result. We should stress that from the exact formulas (\ref{Pin})-(\ref{PinM})  it follows that for any non-zero but arbitrary small  values of parameters, 
$\overline{\mathcal P_{in}} \to 0$ when $\omega \to \infty$. In particular, in the strong damping regime, the mass parameter $M$ is small but non-zero.  So, the reader should take care with the order of limits.  

The similar considerations and methods can be applied to statistical moments for the particle coordinate and velocity as well as to the stationary, time-dependent  averaged values of various forms of energies like (\ref{Ek})-(\ref{Ec}).

\section{Summary}

The Brownian harmonic oscillator is an extremely simple and at the same time paradigmatic model with many applications in both classical and quantum physics. In the paper, we considered a less popular problem of {\it energetics} of the system,  which is permanently moved out of thermal equilibrium, but reaches a stationary state in the long-time regime. Because the model is exactly solved, all interesting characteristics can be analytically evaluated. The first conclusion is that in the stationary state, various forms of averaged energy are additive: one part is related to thermal-equilibrium  energy according to the equipartition theorem and the second part is generated by the external unbiased time-periodic force of frequency $\omega$. Additivity is due to linearity of the system (i.e.,  the potential force is a linear function of the particle coordinate)  and  for non-linear systems it does not hold true.  In the long time limit, when a non-equilibrium stationary state is reached, energy is a periodic function of time of the same period as the external driving. If $\omega > \omega_0$ (where $\omega_0$ is the eigenfrequency of the undamped oscillator), the maximal value of the kinetic  energy is greater than the maximal value of the potential energy.  In turn, if $\omega <  \omega_0$, the maximal value of the kinetic energy is smaller than the maximal value of the potential energy. For  the resonance frequency
$\omega = \omega_0$, the maximal value of the potential energy is equal to the maximal value of the kinetic energy and the total average {\it energy is  conserved}, i.e., it is a constant function of  time. It is the second important result.

We analysed time-independent characteristics of energy  which are obtained by the method of averaging over the period of the time-periodic force. We observe that in some regimes, determined  by  the dissipation strength (quantified  by  the friction coefficient $\gamma$),  the frequency dependence of energy exhibits  non-monotonic behaviour. New results are obtained for the work-fluctuation relation of the Jarzynski type. We analysed this problem in the regime of long times, when the non-equlibrium stationary but time-dependent state is reached. Both the mean value and variance of the work performed on the system  in the interval $(\tau_0, \tau_0 +\tau)$ grows linearly with time $\tau$ (as it should be) but also depend periodically on the reference time $\tau_0$. 
 Finally, we critically discussed the so-called overdamped dynamics. We are of opinion that the correct framework for studying a  regime of strong damping has to be based on analysis of time scales of the system.  The overdamped dynamics cannot simply be obtained by putting the particle mass $m=0$ (which can bring physical nonsense), but rather requires to identify the correct parameters to compute a dimensionless equation, and to calculate the limits for each quantity of interest. It is what we presented in Sec. VII. 
This procedure is always correct and we hope that we were  able to clarify a physically acceptable method of an approximation of some limiting dynamics on a simple example of the driven Brownian harmonic oscillator.

\appendix
\section{Obtaining the Green's function}
\label{apA}

In this appendix we obtain the Green's function of the full Langevin equation. According to the theory of Green's function \cite{mintU}, we first should find the solutions of the corresponding homogeneous differential equation:
\begin{equation}
m\frac{d^2x}{dt^2} + \gamma \frac{dx}{dt} +kx =0,
\end{equation}
The solution takes the exponential form  $y(t)=\mbox{exp}(rt)$,  where $r$ verifies:
\begin{equation}
m r^2 + \gamma r + k = 0. 
\end{equation}
Assuming for simplicity that $\gamma^2 \geq 4mk$, we have two real solutions $r_{+}$ and $r_{-}$ :
\begin{equation}
r_{\pm}=-\frac{\gamma}{2m} \pm \beta, \;\;\;\; \beta=\frac {\gamma}{2m} \sqrt{ 1 - \frac{4km}{\gamma^2}}
\end{equation}
which gives us two independent functions $y_1(t)=\mbox{exp}[(\beta -\gamma/2m)t]$ and $y_2(t)=\mbox{exp}[-(\beta +\gamma/2m)t]$,  respectively. \bigskip

Now, we construct the Green's function $g(t,t')$ that verifies:
\begin{equation}
m\frac{d^2 g(t,t') }{dt^2} + \gamma \frac{d g(t,t')}{dt}  + k g(t,t') = \delta(t-t'),
\label{under}
\end{equation}
with the initial conditions $g(0,t') = 0$ and $d g(t,t')/dt = 0$ for $t=0$ according to the general prescription of finding the Green’s function for the initial value problems.

The Green's function can be written as a linear combination of solutions of the homogeneous equation as follows:
\begin{equation}
  g(t,t')= c_1 y_1 +c_2 y_2, \quad    t<t'
     \end{equation}
and 
\begin{equation}
  g(t,t')=     d_1 y_1 +d_2 y_2, \quad t > t'.
  \end{equation}
We need a set of four equations in order to determine four constants  $c_1$, $c_2$, $d_1$ and $d_2$. The chosen boundary conditions simply gives that $c_1=c_2=0$. 
The continuity of $g(t,t')$ at $t=t'$ implies:
\begin{equation}
c_1 y_1(t') +c_2 y_2(t')=d_1 y_1(t') +d_2 y_2(t')
\end{equation}
which gives
\begin{equation}
d_1 y_1(t') = - d_2 y_2(t')
\label{continuityx}
\end{equation}
The last equation is obtained by integrating equation (\ref{under}) from $t'^+$ to $t'^-$:
\begin{equation}
\int_{t'^-}^{t'^+}\big[ m\frac{d^2 g(t,t') }{dt^2} + \gamma \frac{d g(t,t')}{dt}  + k g(t,t') \big]dt = \int_{t'^-}^{t'^+}\delta(t-t') dt. 
\end{equation}
Because $g(t,t')$ is continuous, therefore  $d g(t,t')/dt$ can only have a jump discontinuity which implies:
\begin{equation}
\frac{dg(t,t')}{dt}\Big|_{t=t'^+}-\frac{dg(t,t')}{dt}\Big|_{t=t'^-}=\frac{1}{m}
\end{equation}
and leads to the relation 
\begin{equation}
\left(-\frac{\gamma}{2m} + \beta\right) d_1 y_1(t')-\left(\frac{\gamma}{2m} + \beta\right) d_2 y_2(t')=\frac{1}{m}
\label{continuityxdot}
\end{equation}
We then use Eqs. (\ref{continuityx}) and (\ref{continuityxdot}) to find:
\begin{equation}
d_2(t')=\frac{-1}{2m\beta y_2(t')}, \;\;\;\;\;\;\;\;\;\; d_1(t')=\frac{1}{2m\beta y_1(t')}
\end{equation}
Finally, the Green's function turns out to be:
\begin{equation}
  g(t,t')= \Theta(t-t') \, 
    \frac{1}{2m\beta} e^{-\frac{\gamma}{2m}(t-t')}\big[ e^{\beta(t-t')} - e^{-\beta(t-t')} \big]  
  \label{greenunder}
\end{equation}
The expression can then be written with the help of Heaviside step function $\Theta(u)$ to retrieve formula~(\ref{Green1}) (and formula~(\ref{Green}) is obtained in the same way if $\gamma^2 \leq 4mk$).

\section{Position and velocity mean square deviation }
\label{appendixB}

In this section, we calculate the mean-squared deviation $\langle (\Delta x^2(t)) \rangle$ and $\langle (\Delta v^2(t)) \rangle$ of the position and velocity of the Brownian partilce, respectively. From (\ref{x(t)}) it can simply be verified that $\langle \Delta x^2(t)\rangle=\langle x^2(t) \rangle - \langle x(t) \rangle^2=\langle x_{\xi}^2(t) \rangle$  and therefore we only need to evaluate the mean-squared of $x_{\xi}(t)$. From (\ref{x3}) we have:
\begin{equation}
x_{\xi}^2(t)=\frac{2\gamma k_B T }{m^2 \Omega^2}\int_0^t \int_0^t dt' dt'' e^{-\gamma_0(2t-t'-t'')} \sin \Omega(t-t') \sin \Omega(t-t'') \xi(t') \xi(t'').
\end{equation}
Taking average over ensemble of noise realizations and using its statistical properties gives:
\begin{equation}
\langle \Delta x^2(t) \rangle=\langle x_{\xi}^2(t) \rangle = \frac{2\gamma k_B T }{m^2 \Omega^2}\int_0^t e^{-2\gamma_0(t-t')} \sin^2 \Omega(t-t')~dt'.
\end{equation}
Evaluation of the integral leads to the following expression for the MSD:
\begin{equation}
\label{msd}
\langle \Delta x^2 (t) \rangle = \frac{k_B T }{k}+\frac{ k_B T }{k \Omega^2} \big[ \gamma_0^2 \cos 2 \Omega t - \gamma_0 \Omega \sin 2 \Omega t - \omega_0^2 \big] \e^{-2 \gamma_0 t}.
\end{equation}
Now, we evaluate the mean-squared velocity deviation: 
\begin{eqnarray}
\label{msdvdef}
\langle \Delta v^2(t)\rangle=\langle v^2(t) \rangle - \langle v(t) \rangle^2=\langle \dot{x}_{\xi}^2(t) \rangle=\langle v_{\xi}^2(t) \rangle, 
\end{eqnarray}
where
\begin{eqnarray}
\label{vxi1}
v_{\xi}(t) =\sqrt{2\gamma k_B T}\,\int_{0}^{\infty} \frac{\partial G(t,u)}{\partial t} \xi(u) du.
\end{eqnarray}
 Differentiation of $G(t, u)$ gives 
\begin{eqnarray}
\label{partialG}
\frac{\partial G(t,u)}{\partial t} &=& \frac{1}{m\Omega}e^{-\gamma_0(t-u)}\bigg[\sin\Omega(t-u)[\delta(t-u)-\gamma_0\Theta (t-u)] \nonumber\\
&+&\Omega\cos\Omega(t-u)\Theta (t-u)\bigg].
\end{eqnarray}
Replacing (\ref{partialG}) into Eq.(\ref{vxi1}) leads to the relation 
\begin{equation}
\label{vxi2}
v_{\xi}(t) =\frac{\sqrt{2\gamma k_B T}}{m\Omega}\int_{0}^t e^{-\gamma_0(t-u)} [\Omega\cos \Omega(t-u)-\gamma_0\sin \Omega(t-u)] \xi(u) du,
\end{equation}
and it follows:
\begin{equation}
\langle \Delta v^2(t) \rangle=\langle v_{\xi}^2(t) \rangle = \frac{2\gamma k_B T }{m^2 \Omega^2}\int_0^t du e^{-2\gamma_0(t-u)} [\Omega\cos \Omega(t-u)-\gamma_0\sin \Omega(t-u)]^2.
\end{equation}
The integration yields:
\begin{equation}
\label{msdv}
\langle \Delta v^2(t) \rangle=\frac{ k_B T }{m } + \frac{ k_B T }{m \Omega^2} (\gamma_0\Omega\sin 2\Omega t - 2 \gamma_0^2 \sin^2 \Omega t - \Omega^2) e^{-2\gamma_0t}.
\end{equation}

\section{Statistical moments of the work}

In the long-time regime, when $\gamma_0 \tau_0 >> 1$, the mean value of the work is
\begin{eqnarray}
\label{W1}
\mu_{\tau_0}(\tau) =  \omega F_0 \int_{\tau_0}^{\tau_0+\tau} 
\sin(\omega t) \langle x(t)\rangle  dt  = \omega F_0 \int_{\tau_0}^{\tau_0+\tau} \sin(\omega t) X_d(t) dt   \nonumber \\  
= \overline{\mathcal P_{in}} \; \tau - \frac{1}{2\omega} \overline{\mathcal P_{in}} \left\{\sin[2\omega (\tau_0 +\tau)] - \sin(2\omega \tau_0)\right\}
\nonumber\\
+ \frac{\omega^2 -\omega_0^2}{2m A(\omega)} \left[F^2(\tau_0 +\tau) - F^2(\tau_0)\right],  
\end{eqnarray}
where  $\langle x(t)\rangle =X_d(t)$ is given by equation (14),  $F(t)=F_0 \cos(\omega t)$ is the external time-periodic force,  $A(\omega)$ is defined in equation (13) and the period-averaged input power $\overline{\mathcal P_{in}}$ is determined by equation (45). We have to emphasize that in the long-time stationary state the mean work still depends on the reference time $\tau_0$:  If $\tau$ is fixed and if one translate the reference  time $\tau_0 \to \tau_0 + \epsilon$ then in general the mean work will be different. 
However, the mean work performed over one period  $\tau=\mathsf{T} =2\pi/\omega$  of the driving $F(t)$ does not depend on $\tau_0$ and its value is 
$\mu_{\tau_0}(\mathsf{T})  = \overline{\mathcal P_{in}} \; \mathsf{T}$. It is in agreement with results discussed in section 6. 
 
The second statistical moment of the work  is  
\begin{equation}
\label{W2}
\langle W^2_{\tau_0}(\tau) \rangle =  \omega^2 F_0^2  \int_{\tau_0}^{\tau_0+\tau} dt  \sin(\omega t) \int_{\tau_0}^{\tau_0+\tau} du 
 \sin(\omega u)  
 \langle x(t) x(u)\rangle.  
\end{equation}
where $x(t)$ is given by equation (5). 
The correlation function of the particle position is 
\begin{equation}
\label{xtxu}
 \langle x(t) x(u)\rangle  = x_d(t) x_d(u) + 2\gamma k_B T \int_0^t ds  \int_0^u dr \; G(t, s) G(u, r) \langle \xi(s) \xi(r)\rangle,    
\end{equation}
where the Green function $G(t, s)$ is given by equation (9). 
We have neglected the exponentially decaying term $x_c(t)$ which includes the initial conditions, see equation  (6). 
   The correlation function of thermal noise is 
$\langle \xi(s)\xi(r)\rangle =\delta(s-r)$. 
Because the variable $t$ and $u$ take values in the same interval, the cases 
$t>u$ and $t<u$  should be considered:  
\begin{eqnarray}
\label{Xtxu}
 \langle x(t) x(u)\rangle  = x_d(t) x_d(u) + 2\gamma k_B T \; \theta(t-u) 
 \int_0^u ds  \; G(t,s) G(u,s) \nonumber\\
 + 2\gamma k_B T \;\theta(u-t) 
 \int_0^t ds \;  G(t, s) G(u,s).  
\end{eqnarray}
We insert it into eqution ({\ref{W2}) and get  
\begin{eqnarray}
\label{WW2}
 \langle W^2_{\tau_0}(\tau) \rangle  = 
 \langle W_{\tau_0}(\tau)\rangle^2
 \nonumber\\ 
 + 4\gamma k_B T  \omega^2 F_0^2 \;  \int_{\tau_0}^{\tau_0+\tau} dt \sin(\omega t)  \int_{\tau_0}^{t} du 
  \sin(\omega u) \int_0^u ds  \; G(t,s) G(u,s). 
  \end{eqnarray}
We have used symmetry of the correlation function $\langle x(t)x(u)\rangle$ under the interchange of time arguments. In consequence, two terms in r.h.s of equation ({\ref{xtxu}) give the same contribution to the second moment of the work. 
From the above equation it follows that the variance of the work is 
\begin{eqnarray}
\label{variance}
 \sigma^2_{\tau_0}(\tau) 
 = \frac{4\gamma k_B T  \omega^2 F_0^2}{m^2 \Omega^2} \; \nonumber \\
 \times \int_{\tau_0}^{\tau_0+\tau} dt  e^{-\gamma_0 t} \sin(\omega t)  \int_{\tau_0}^{t} du e^{-\gamma_0 u}
  \sin(\omega u) \int_0^u ds  \; e^{2\gamma_0 s} \sin[\Omega(t-s)]\nonumber \\
  \times   \sin[\Omega(u-s)].   
  \end{eqnarray}
It is convenient to express all trigonometric functions by their exponential forms. Then integrals of exponential functions can easily  be calculated. The only cumbersome task is to simplify coefficients in the final expressions. One could use computer algebra to ease this part. The final result reads
\begin{eqnarray}
\label{varsigma}
\beta  \sigma^2_{\tau_0}(\tau) 
 = 2 \overline{\mathcal P_{in}} \; \tau \nonumber\\
 - \overline{\mathcal P_{in}} 
 \left\{ \frac{1}{\omega}\left[\sin\left(\omega(\tau_0+\tau)\right) -\sin(\omega \tau_0)\right] 
 - \tau_1 C_1 \left[\cos\left(\omega(\tau_0+\tau)\right) -\cos(\omega \tau_0)\right]\right\} \nonumber\\
 + \overline{\mathcal P_{in}} \,\tau_1  \left[ e^{-\gamma_0 \tau} I(\tau_0, \tau) - I(\tau_0, 0)\right], 
  \end{eqnarray}
where the characteristic (relaxation) time $\tau_1=m/\gamma =1/2\gamma_0$ and the periodic function $I(\tau_0, \tau)$ has the following structure 
\begin{eqnarray}
\label{I}
I(\tau_0, \tau) = \alpha_0 \cos\left[2\omega \tau_0+ \omega \tau)\right] \left[
\alpha_1  \cos(\Omega \tau) + \alpha_2 \frac{ \sin(\Omega \tau)}{\Omega}\right] \nonumber\\
+ \cos(\Omega \tau)\left[ G_1 \cos(\omega \tau) + G_2 \sin(\omega \tau)\right] 
+ \frac{\sin(\Omega \tau)}{\Omega} \left[ G_3 \cos(\omega \tau) + G_4 \sin(\omega \tau)\right].  \nonumber\\
\ \ \ 
  \end{eqnarray}
All dimensionless parameters $C_1, \alpha_k (k=0, 1, 2)$ and $G_n (n=1, 2, 3, 4)$ are expressed by three  system parameters $\{\gamma_0, \omega_0, \omega\}$. We do not present their explicit form because we want to discusse a fundamental issue and not details.  

From equation (\ref{varsigma}) we infer that fluctuations of the work, quantified by its variance,   depend on the reference time $\tau_0$ and  consist of several characteristic contributions. The  part $ 2 \overline{\mathcal P_{in}} \; \tau $ 
grows linearly with the time-interval $\tau$. It would correspond to normal diffusion, not in a position space but  in the work space.  There is  a part  periodic in $\tau$ and a contribution  which exponentially decays as $\tau$ increases. There is also a part  independent of $\tau$ (in the term $I(\tau_0, 0)$).  

\section*{Acknowledgements}

This work was supported by the NCN grant 2015/19/B/ST2/02856 (J{\L}). We would like to express our gratitude to Artem Ryabov, Viktor Holubec and Peter Chvosta from Charles university in Prague. MEF acknowledges useful discussion with Ali Najafi from IASBS in Zanjan. J{\L} thanks Chris Jarzynski for discussion on the work-fluctuation  relations.

\end{document}